\documentclass[usenatbib]{mn2e}
 \usepackage[dvips]{graphicx}   
\usepackage{amsmath}
\usepackage{epsfig}
\def\lsim{\mathrel{\mathop
  {\hbox{\lower0.5ex\hbox{$\sim$}\kern-0.8em\lower-0.7ex\hbox{$<$}}}}}
\def\gsim{\mathrel{\mathop
  {\hbox{\lower0.5ex\hbox{$\sim$}\kern-0.8em\lower-0.7ex\hbox{$>$}}}}}

\def\lapprox{\hbox{\lower .8ex\hbox{$\,\buildrel < \over\sim\,$}}}
\def\gapprox{\hbox{\lower .8ex\hbox{$\,\buildrel > \over\sim\,$}}}

\def\thetaB{\mbox{\boldmath$\theta$}}

\def\xisp{\xi(\sigma, \pi)}

\def\xir{\xi(r)}
\def\xips{\xi(\pi,\sigma)}

\newcommand{\nc}{\newcommand}

\nc{\be}[1]{\begin{equation}\mbox{$\label{#1}$}}
\nc{\bea}[1]{\begin{eqnarray} \mbox{$\label{#1}$}}
\nc{\Section}[2]{\section{#2}\label{#1}}
\nc{\Bibitem}[1]{\bibitem{#1}}
\nc{\Label}[1]{\label{#1}}

\nc{\Mpc}{Mpc/h}
\nc{\vev}[1]{\langle #1 \rangle}

\nc{\eea}{\end{eqnarray}}
\nc{\ee}{\end{equation}}

\begin{document}


\title[Baryon Acoustic Peak in the Line-of-Sight Direction]
{Clustering of Luminous Red Galaxies IV:\\
Baryon Acoustic Peak in the Line-of-Sight Direction\\
and a Direct Measurement of $H(z)$}

\author[E.Gaztanaga, A.Cabre, L.Hui]{Enrique Gazta\~naga$^{1}$, 
Anna Cabr\'e$^{1}$ \& Lam Hui$^{2}$\\
$^{1}$Institut de Ci\`encies de l'Espai, IEEC-CSIC, 
F. de Ci\`encies, Torre C5 par-2,  Barcelona 08193, Spain\\
$^{2}$Institute for Strings, Cosmology and Particle Physics, Columbia Astrophysics Laboratory,
and Department of Physics, \\
Columbia University, New York, NY 10027, U. S. A.}

\twocolumn
\maketitle

\begin{abstract}
We study the clustering of LRG galaxies in the latest spectroscopic SDSS 
data releases, DR6 and DR7, which sample over 1 ${\,\rm Gpc^3/h^3}$ to z=0.47. 
The 2-point correlation function $\xisp$ is estimated as a function of
perpendicular $\sigma$ and line-of-sight $\pi$ (radial) directions.
We find a significant detection of a peak at $r\simeq 110$Mpc/h,
which shows as a circular ring in the $\sigma-\pi$ plane.
There is also significant evidence for a peak along the radial direction
whose shape is consistent with its originating
from the recombination-epoch baryon acoustic oscillations (BAO).
A   $\xisp$ model with no radial BAO peak is disfavored 
at $3.2\sigma$, whereas a model with no magnification bias is disfavored at $2\sigma$.
The radial data enable, for the first time, a direct measurement
of the Hubble parameter $H(z)$ as a function of redshift. 
This is independent from earlier BAO measurements which used the spherically averaged
(monopole) correlation to constrain an integral of $H(z)$.
Using the BAO peak position as a standard ruler in the radial direction, we find: 
$H(z=0.24)= 79.69 \pm 2.32 (\pm 1.29)$ 
km/s/Mpc for z=0.15-0.30 and $H(z=0.43)= 86.45 \pm 3.27 (\pm 1.69)$ km/s/Mpc 
for $z=0.40-0.47$. The first error is a model independent 
statistical estimation and the second accounts for systematics
both in the measurements and in the model. For the full sample, $z=0.15-0.47$, we find
$H(z=0.34)= 83.80 \pm 2.96 (\pm 1.59)$ km/s/Mpc. 

\end{abstract}

\maketitle

\section{Introduction}
\label{intro}

Luminous red galaxies (LRG's) are selected by color and magnitude to obtain intrinsically 
red galaxies in SDSS (Eisenstein etal 2001). 
These galaxies trace a big volume, 
around $1 {\rm \, Gpc^3 h^{-3}}$, which makes them ideal for studying clustering
on large scales \citep[see][]{Hogg2005}. Attention has been paid especially to
the baryon acoustic peak around a scale of $100$ Mpc/h, because of its value
as a standard ruler. In Eisenstein etal (2005)
the baryon acoustic peak was detected in the spherically averaged
two-point correlation function (i.e. the monopole) using LRG's from an earlier
SDSS release (about half the size of the final data). Both
2dFGRS and SDSS spectroscopic redshift surveys have been used
to constrain cosmological parameters 
via the galaxy power spectrum \citep{clusteringlrg,sanchez}, including information
from the baryon acoustic feature \citep{hutsia, hutsib, percival, Sanchez09}. 
Photometric LRG catalogs cover a larger volume and larger densities and have also
been used to obtain cosmological constraints \citep{padmanabhan2007,blake}.

In this paper, as in Papers I \citep{paper1} and II \citep{paper2} of this series,
we focus on the LRG's anisotropic
redshift space correlation function $\xi(\sigma, \pi)$, where $\pi$ is the
line-of-sight (LOS) or radial separation and $\sigma$ is the transverse separation.
There are three sources of anisotropy at scales larger than 40Mpc/h. 
It is well known that peculiar motion distorts the
correlation function anisotropically \citep{Kaiser}. It is also well known that
assuming the wrong background cosmology will lead to an anisotropic $\xi$ \citep{AP}.
What is under-appreciated is that gravitational lensing also introduces an
anisotropy of its own to the galaxy correlation function \citep{matsulen,hui1,hui2}.
In this paper, we will present evidence of this lensing distortion in the LOS direction.

A distinct focus of this paper in the series is on the baryon acoustic oscillation (BAO) feature
in $\sigma-\pi$ plane.  In principle,
it can be used as a standard ruler to measure both the Hubble expansion rate $H(z)$,
via its observed redshift span in the radial direction, and the 
angular diameter distance $D_A (z)$,
via its observed angular size in the transverse direction \citep{glazebrook,seoeisenstein}.
We will implement this idea to measure $H(z)$. We are aided in this endeavor by
two effects. Redshift distortions in the LOS direction
make the $\xi$ negative on intermediate scales ($\sim 50 - 90$ Mpc/h), while
magnification bias gives $\xi$ a positive boost especially on large scales ($\gsim 100$ Mpc/h).
This combination  enhanced the contrast of the radial BAO peak with respect to the 
BAO in the perpendicular direction.
Because the noise is shot-noise dominated any increase in the signal, such as bias
or magnification bias, increase the signal-to-noise and helps the BAO detection.
We will show that the shape of the correlation function in the radial direction is
in good agreement with the predictions.
We will also validate the interpretation of the feature as baryon acoustic in nature
by studying its appearance in directions away from the radial direction,
and performing a parametric fit to the monopole.

It is useful to point out some related earlier work.
Redshift distortions in the LRG's and quasars at $z\sim 0.55$ have been studied
using the 2dF-SDSS LRG and QSO Survey \citep[2SLAQ][]{cannon,ross,daangela,wake}. 
\citet{okumura} measured $\xisp$ away from the LOS direction
in the SDSS using a sample similar
to ours: they used about 47000 LRG's over a redshift range of $0.16 - 0.47$
while we use 75000 LRG's over $z$ of $0.15 - 0.47$.
None of the earlier papers attempted a direct measurement of $H(z)$ from
the baryon feature in the LOS direction.

This paper is organized as follows. Section \ref{sec:theory} gives a summary of the theory, 
including a brief introduction to magnification bias. In section \ref{sec:analysis} 
we perform an analysis of the clustering signal around the position of the BAO peak, 
both in the monopole and in the $\sigma-\pi$ plane, especially in the radial direction, 
from which we obtain the Hubble parameter H(z) in section \ref{sec:interpretation}. 
We end with a conclusion in section \ref{sec:conclusions}, where we deduce implications 
for the dark energy equation of state, and  where we emphasize the need for further 
theoretical work on the nonlinear coupling between peculiar motion
and magnification bias.

\section{Theory}
\label{sec:theory}

The 2-point correlation function, $\xi(\vec{r})$, is defined by the joint
probability that two galaxies are found in two volume elements 
$dV_1$ and $dV_2$ placed at separation $\vec{r}$ \citep[see][]{peebles1980}:

\begin{equation}
 dP_{12}=n^2[1+\xir]dV_1dV_2
\label{eq:xir}
\end{equation}
where $n$ is the mean number density of galaxies.

\begin{figure}
\centering{ \epsfysize=7.cm\epsfbox{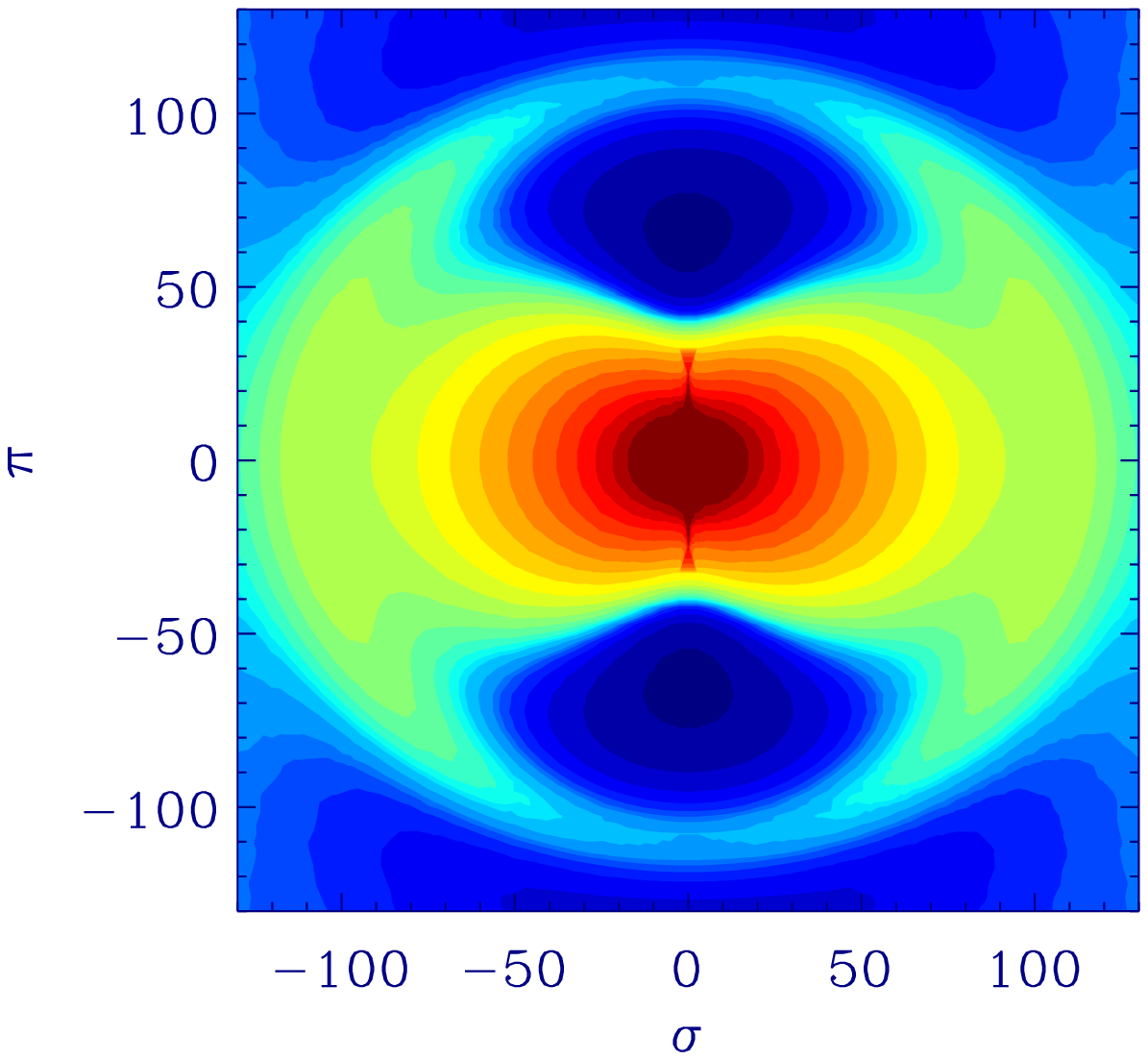}}
\centering{ \epsfysize=7.cm\epsfbox{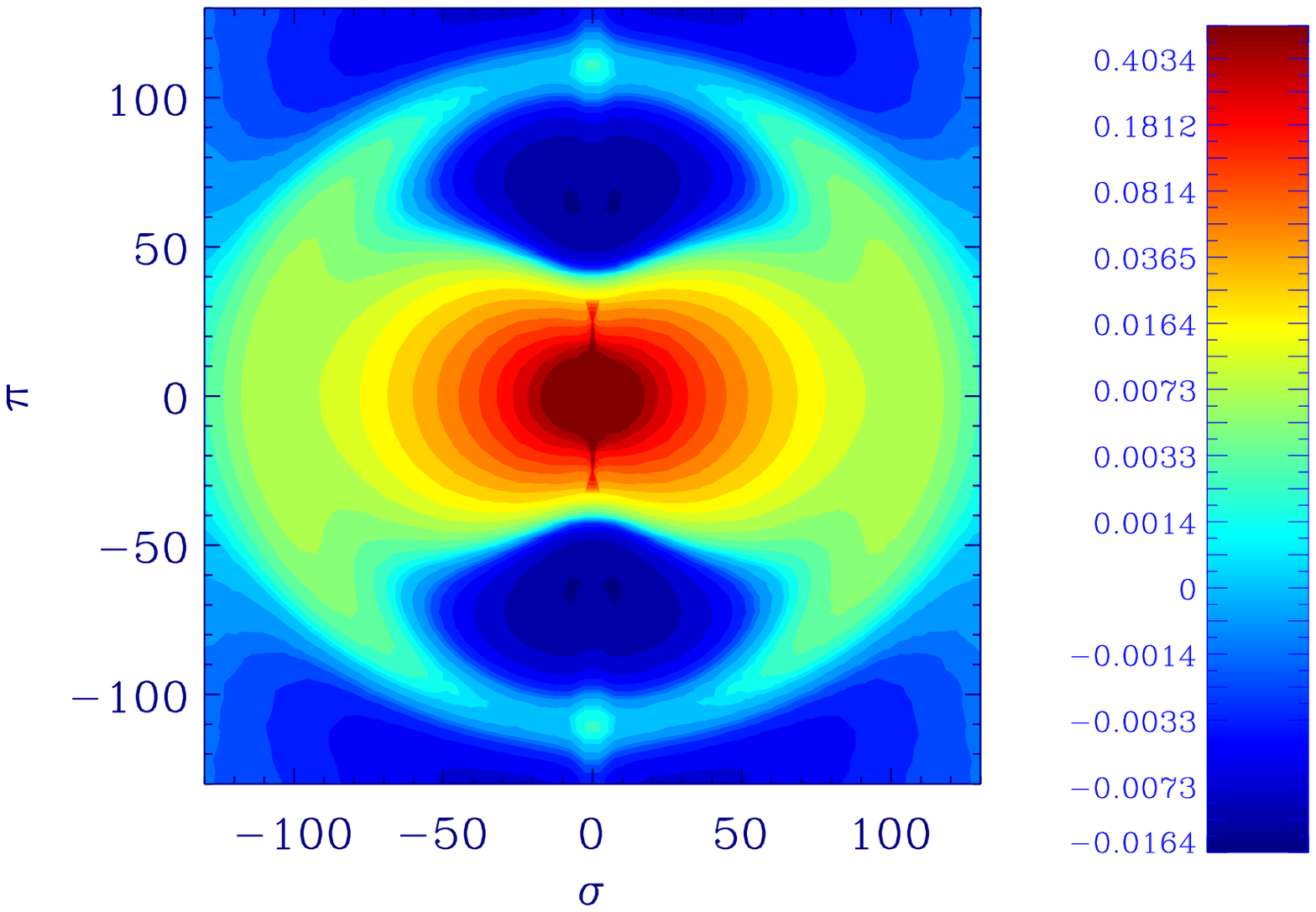}}
  \caption{{\it Top panel:} 
Theoretical $\xisp$ with linear redshift space distortions convolved
with a dispersion model
, assuming a cosmology and bias (linear and non-linear) as observed in LRG in paper I and II.
Note the 
ring around $\simeq 100$ Mpc/h that becomes narrow and
less prominent in the radial direction. 
{\it Bottom panel:} same with magnification bias added (slope = 2 to see clearly the effect). The main effect
is the boosting of the BAO peak in the $\pi$ direction. In both cases we take into account the bin smoothing at 5Mpc/h.
\label{fig:baoteoric1} }
\end{figure}


\begin{figure*}
\centering{ 
\epsfysize=10.cm\epsfbox{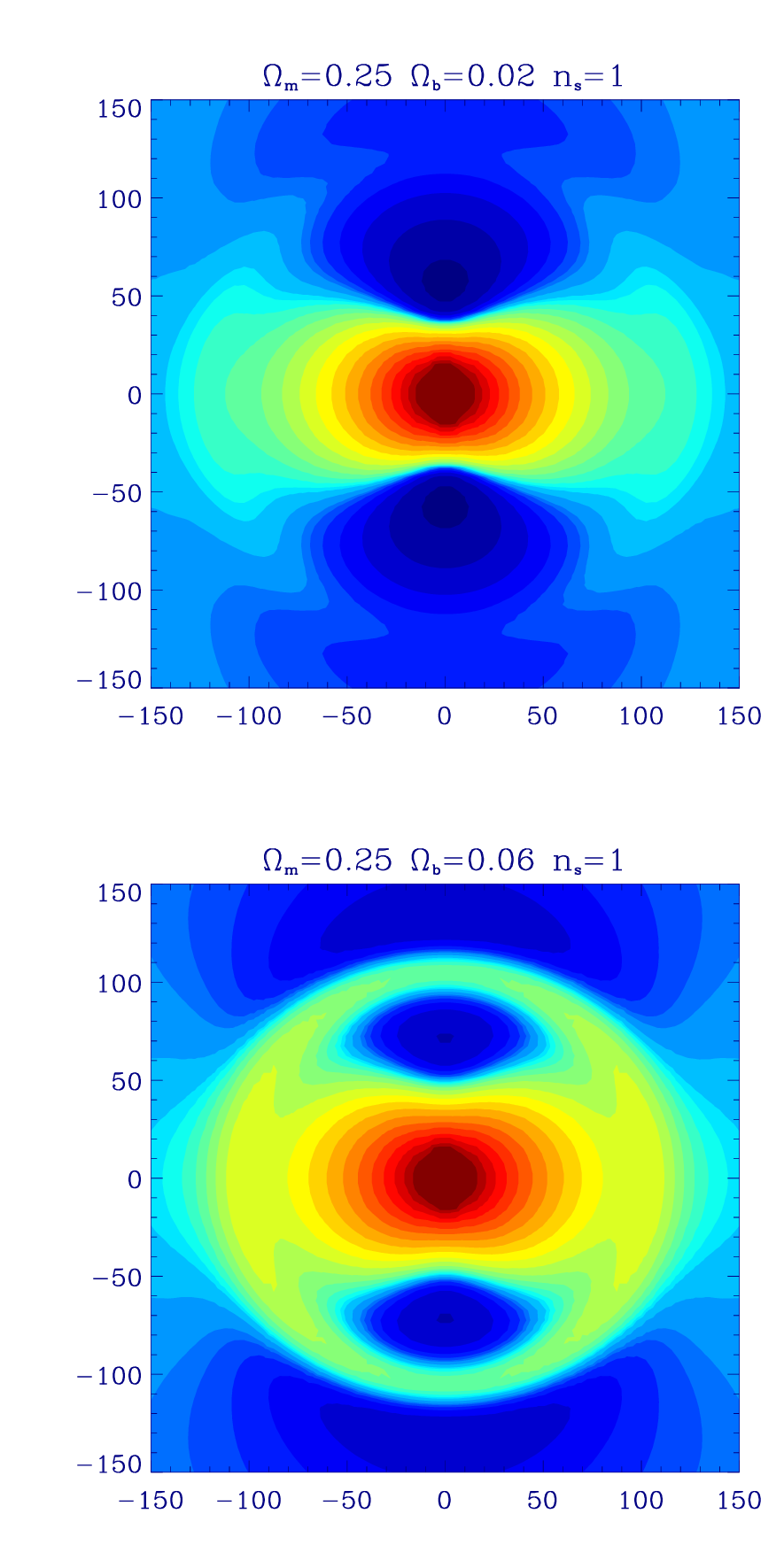}}
\centering{ 
\epsfysize=10.cm\epsfbox{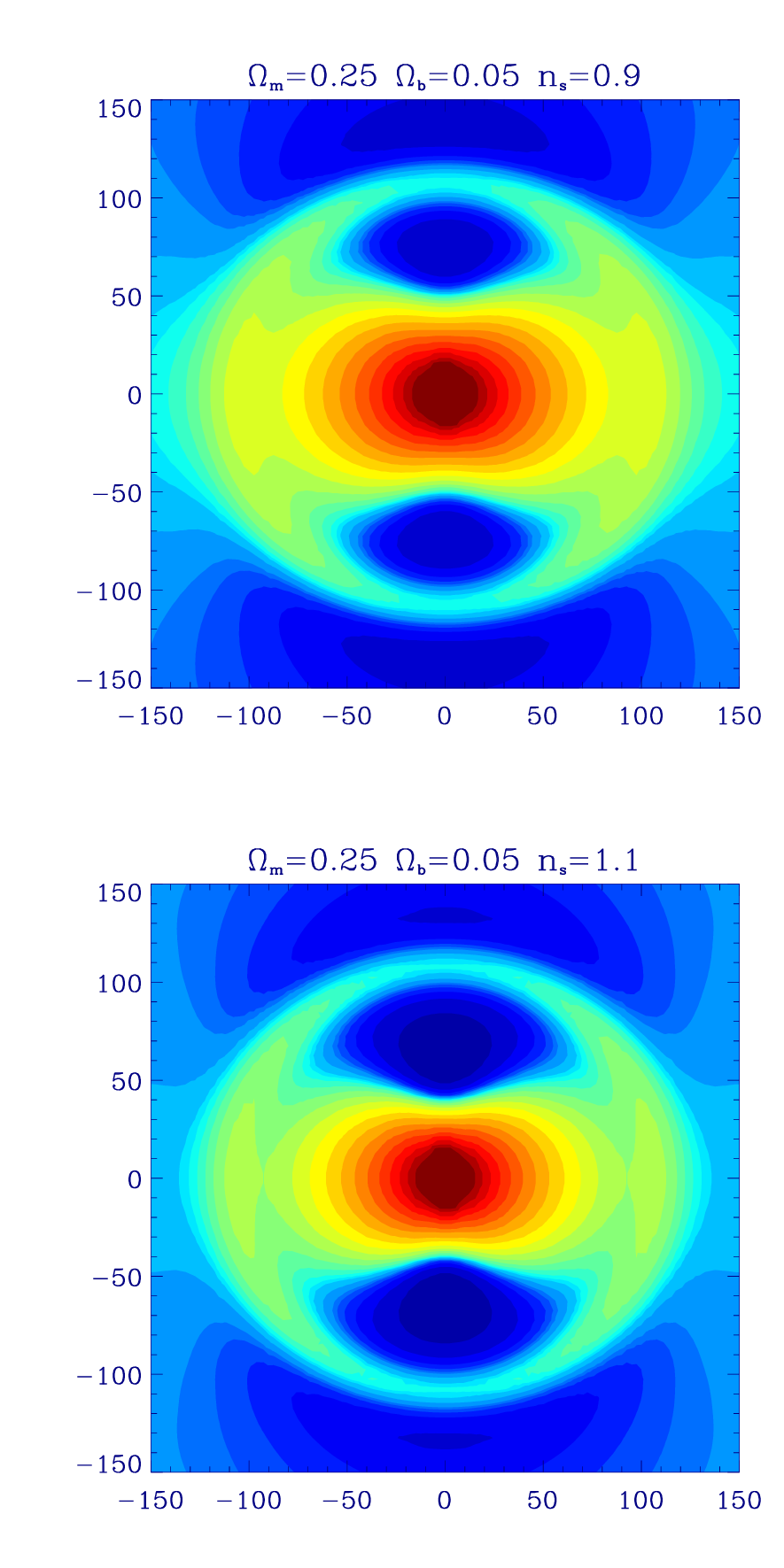}}
\centering{ 
\epsfysize=9.7cm\epsfbox{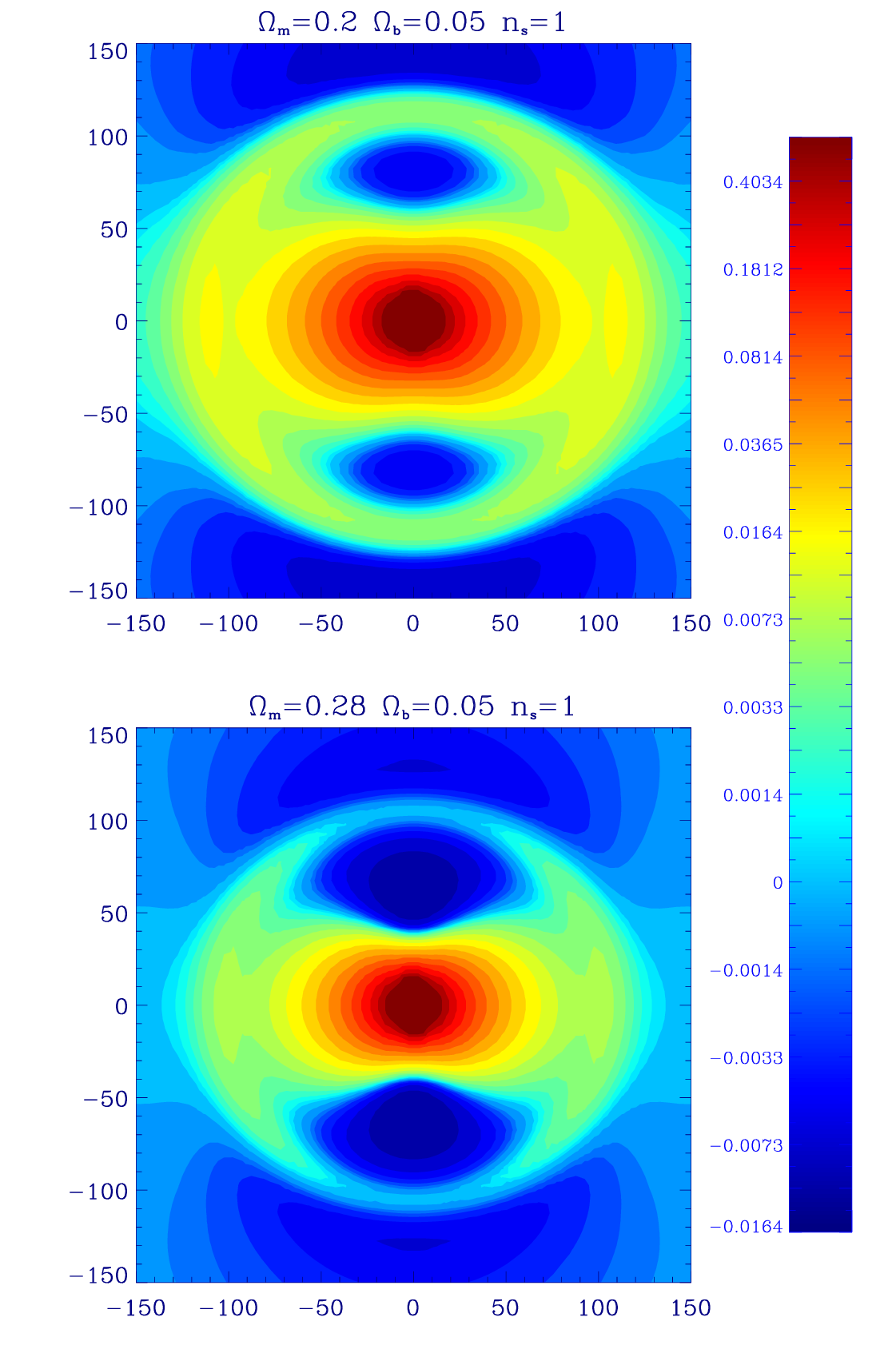}}
  \caption{Same as Fig.\ref{fig:baoteoric1}, with
different panels corresponding to different cosmological parameters as labeled
in the figures: left panels show change with the baryon density $\Omega_b$, middle panels
with the scalar spectral index $n_s$ and right panels with the matter density $\Omega_m$.
\label{fig:baoteoric} }
\end{figure*}

We can split the distance $\vec{r}$ into its component along the
line-of-sight
(LOS) $\pi$ and perpendicular to the LOS $\sigma$,
where $r^2 = \pi^2 + \sigma^2$.
Azimuthal symmetry implies $\xi$ is in general a function of $\pi$ and $\sigma$ alone.

The correlation $\xisp$ is related to the power
spectrum by a Fourier transform:

\begin{equation}
 \xisp=\int P_s(\vec{k})e^{-i\vec{k}\vec{r}}\frac{d^3k}{(2\pi)^3}
\label{eq:fourier}
\end{equation}

Note that the use of $\pi$ to denote the LOS separation is conventional.
Hopefully the reader will not confuse it with the numerical constant $\pi$ as
in $(2\pi)^3$.

\subsection{Redshift Distortions by Peculiar Motion}
\label{zdistortion}

In the large-scale linear regime, and in the plane-parallel approximation 
(where galaxies are taken to be sufficiently far away from the observer 
that the displacements induced by peculiar velocities are effectively parallel),
the distortion caused by coherent infall velocities takes a particularly 
simple form in Fourier space
\citep{Kaiser}:

\begin{equation}
P_s(\vec{k}) = (1 + \beta\mu_k^2)^2 P_{gg} (k).
\label{eq:kaiserpoint}
\end{equation}
where $P_{gg} (k)$ is the power spectrum of galaxy density fluctuation $\delta_g$,
$\mu$ is the cosine of the angle between $\vec{k}$ and the line-of-sight, the subscript $s$ indicates 
redshift space, and $\beta$ is the growth rate of growing modes in linear theory,
the dimensionless quantity which solves the linearized continuity equation
${\vec{\nabla}} . {\vec{ v}} + (a'/a) \beta \delta_g = 0$, where
the prime denotes derivative with respect to conformal time.
Assuming that the galaxy over-density $\delta$
is linearly biased by a factor $b$
relative to the underlying matter density $\delta_{m}$, i.e.
$\delta_g = b \delta_m$, and that the velocities are unbiased, the value of $\beta$
can be approximated by
\begin{equation} 
 \beta \approx {\Omega_m^{0.55} \over b}
\label{fb}
\end{equation}
\citep[see][for a review]{hamilton1992}. After integration in Eq.\ref{eq:fourier}, 
these linear distortions in $P_s (\vec{k})$ produce a distinctively anisotropic $\xisp$.
Redshift distortions in the linear regime  produce a lower amplitude and sharper
 baryon acoustic peak in the LOS 
than in the perpendicular direction because 
of the coherent infall into large scale overdensities.
This is illustrated in top panel of Fig.\ref{fig:baoteoric1}. A characteristic
feature of this effect is a valley of negative correlations (in blue)
on scales between $\pi = 50-90$ Mpc/h, which as we will show is in
good agreement with our measurements from real data.
Such a valley is absent without redshift distortions.

Redshift space distortions on smaller scales (commonly called the finger-of-god effect)
are often approximated by a dispersion model where $\xi$ is convolved with a
distribution of pairwise velocities, parametrized by a single parameter $\sigma_v$.
Details are presented in Paper II \citep{paper2}. 
Note that in Fig. \ref{fig:baoteoric1}, as in all $\xi(\sigma, \pi)$ contour diagrams
of this paper, bins of $5$ Mpc/h $\times$ $5$ Mpc/h are used.
The finger-of-god effect is much suppressed by using bins of this size.
Readers interested in an analysis that brings out this effect rather than
suppressing it are referred to Paper II \citep{paper2}.

Fig.\ref{fig:baoteoric} shows how the shape of $\xips$ changes with cosmological
parameters: baryon density $\Omega_b$,
initial spectra index $n_s$ and cold dark matter density $\Omega_m$. 
Note how the BAO ring becomes more or less asymmetric and how this
is correlated with the valley of negative correlations (in blue).

\subsection{Magnification bias}

\begin{figure}
\centering{ \epsfysize=5.cm\epsfbox{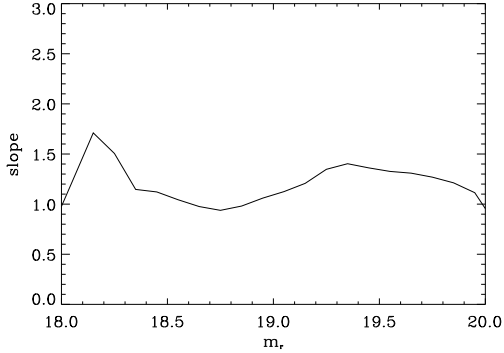}}
\caption{LRG 
 number count slope $s$ 
as a function of limiting apparent magnitude $m_r$ using all SDSS DR6 photometric catalog.
\label{fig:slopef}}
\end{figure}

Gravitational lensing inevitably modulates the observed
spatial distribution of galaxies. To lowest order, there
are two effects at work. Imagine a number of galaxies located behind
some large mass concentration. Dim galaxies that otherwise would not
have been detected are brought into one's sample by the lensing magnification.
This increases the observed number density of galaxies.
On the other hand, magnification also increases the apparent area, which
leads to a drop in the observed number density of galaxies.
The net lensing effect, known as magnification bias,
is controlled by the slope of the number counts
\citep{tog,webster,fugmann,narayan,schneider,btp,moessner}:
\begin{eqnarray}\label{eq:slopedef}
s = {d {\,\rm log}_{10} N(< m) \over dm}
\end{eqnarray}
where $N(<m)$ refers to the number of galaxies in the survey with apparent magnitude brighter
than $\;m$. Note that to estimate $s$ we need to know $N(<m)$ for magnitudes that are
fainter that our spectroscopic limit of $m=19.2$. We can use the parent photometric DR6 sample,
which is deeper but does not have redshifts.
 Fig. \ref{fig:slopef} shows the estimation for $s$ in the SDSS LRG DR6 galaxies extracted 
from the full (parent) photometric catalog. Because the parent catalog is deeper, it also goes further in redshift.
Thus a fraction of the fainter galaxies in the photometric catalog will be at different redshift to the 
galaxies in our spectroscopic redshift samples. 
Also note that different selection effects go into the photometric and spectroscopic
samples. Thus the true number count slope in the spectroscopic sample is somewhat uncertain
and might be different for different redshift slices\footnote{Note that the
number counts here are limited to some particular redshift range, e.g. z=0.15-0.30,
and should not be confused with the integrated number counts from redshift zero.
In other words, the slope of the number count $s$ here should best be thought
of as being related to the slope of the intrinsic luminosity function.
}. 
Our magnitude cut is around $19.2$, where $s$ seems to lie close to $1.5$, with 
a potentially large uncertainty if evolutionary and/or selection effects turn out to be important.
Given that variations of $s$ in Fig.\ref{fig:slopef} are of order $\Delta s\simeq 0.5$ we consider
a range around $s=1.0-2.0$ for our modeling of magnification bias.

The observed galaxy overdensity $\delta_{obs}$ is related to the
intrinsic (or true) galaxy overdensity $\delta_g$ by:
\begin{eqnarray}s
\label{dobs}
\delta_{obs} = \delta_g + \delta_\mu
\end{eqnarray}
where $\delta_\mu = (5s-2) \kappa$, with $\kappa$ being the lensing convergence
which is simply a weighted LOS integral of the mass fluctuation $\delta_m$
\citep[see][for details]{hui1}

Including linear redshift distortions by peculiar motion is straightforward:
\begin{eqnarray}
\delta_{obs} = \delta_g + \delta_\mu + \delta_v
\label{eq:deltaobs}
\end{eqnarray}
where $\delta_v = (1 + \bar z) H(\bar z)^{-1} \partial v_\pi/\partial \pi$ with
$\bar z$ being the mean redshift, and $v_\pi$ being the peculiar velocity in the $\pi$ direction.
Squaring the above expression, the net observed correlation function is therefore
\begin{equation}\label{eq:observed}
\xi_{obs}=\xi_{gg}+\xi_{gv}+\xi_{vg}+\xi_{vv}+\xi_{g\mu}+\xi_{\mu g}+\xi_{\mu\mu}
\end{equation}
The velocity-magnification cross-terms are absent by virtue of Limber approximation and linear
theory \citep[see][for details]{matsulen,hui2}.
The first term is the true or intrinsic galaxy clustering signal.
The next three terms account for the Kaiser effect. These four terms
together are the real space analog of what is shown in Eq.(\ref{eq:kaiserpoint}).
The rest of the terms account for magnification bias: the magnification-magnification
term is very small at our moderate redshifts of interest; we need focus only on
the galaxy-magnification cross terms \citep{hui1,hui2}.
They are capable of altering significantly the observed correlation function on large scales.
Taylor expanding in $\pi/\bar \chi$, where $\pi = |\chi_1 - \chi_2|$ is the LOS separation
between a pair of galaxies and $\bar \chi$ is the mean sample depth, the galaxy-magnification
cross correlation can be written as \citep{hui1}:
\begin{eqnarray}
\label{eq:xigmu}
\xi_{g\mu} (\chi_1, \thetaB_1; \chi_2, \thetaB_2) + \xi_{\mu g} (\chi_1, \thetaB_1; \chi_2, \thetaB_2) =
\\ \nonumber 
{3\over 2} H_0^2 \Omega_m (5s - 2) (1+\bar z)
| \chi_1 - \chi_2 | \\ \nonumber
\int {d^2 k_\perp \over (2\pi)^2} P_{gm} (\bar z, k_\perp) 
e^{i {\bf k_\perp} \cdot \bar\chi (\thetaB_1 - \thetaB_2)}
\end{eqnarray}
where $P_{gm}$ is the galaxy-mass power spectrum.
Here, $H_0$ is the Hubble constant today, $\Omega_m$ is the matter density,
$\bar z$ is the mean redshift of the sample in question, and
$\chi$ and $\thetaB$ with subscripts label the radial distance and angles of
a pair of galaxies.

The above expression shows clearly the anisotropic nature of
the lensing corrections to the galaxy correlation function.
A large LOS separation $\pi = |\chi_1 - \chi_2|$ is clearly favored as it should be,
since lensing is more effective if the source (background galaxy) and the lens (foreground galaxy)
are further apart. On the other hand, a small transverse separation $\sigma \sim \bar\chi
|\thetaB_1 - \thetaB_2|$ is helpful for boosting the lensing correction. This makes sense
because gravitational lensing is strongest when the impact parameter is small.
The magnification distortion of the correlation function hence is strongest in the LOS $\pi$
direction. Note that the intrinsic galaxy clustering strength generally drops
with separation, even as the lensing correction increases with the LOS separation $\pi$. This means
for a sufficiently large LOS separation, the lensing correction could dominate.

It is also important to note that for a small transverse separation $\sigma$,
the integral of $P_{gm}$ is dominated by high wavenumbers, including ones where
both the mass fluctuations and the galaxy bias are nonlinear.
These nonlinear effects can further boost the galaxy-magnification correlation.
Indeed, if $\sigma$ were small enough, one might need to worry about strong lensing effects.
But since throughout our analysis, we avoid any pairs with $\sigma$ less than $0.5$ Mpc/h
(because of fiber collision issues), we are safely in the weak lensing regime.

The bottom panel of Fig.\ref{fig:baoteoric1}
shows the theoretical expectation for $\xisp$
with magnification bias taken into account.
This can be compared against the top panel, which has no magnification bias.
There is a clear enhancement of the BAO in the radial direction ($\sigma = 0$).
The effect in other directions is relatively minor.
Here, we have used a small scale galaxy bias similar to that in LRG galaxies (see Fig.\ref{fig:biasnl}) and a number count slope
of $s = 2$. The large scale galaxy bias is $2$.
See \S \ref{radialpeak}, \citet{jss} and \citet{hui1} for further discussions on
these numbers. 

\subsection{Beyond Standard Theory}
\label{sec:nonstandard}

The above discussion of the theory for redshift and magnification
distortions is more or less standard (though
the anisotropy induced by lensing is not as widely appreciated).
There are, however, certain limitations to the standard theory.
For instance, the mapping from real to redshift space creates nonlinear effects
even on very large scales. A sign of this is that caustics can occur
in redshift space even if the real space overdensity is rather modest \citep{hks}. 
As emphasized by \citet{scoccimarro}, depending on the statistics studied,
the Kaiser plus dispersion model for redshift distortions
can be inaccurate even on linear scales. 
Similar concerns apply to magnification distortions when peculiar motion
are taken into account. The expression in Eq. \ref{eq:observed} ignores
velocity-magnification cross-correlation. This is justified in linear theory,
but not valid once nonlinear effects are taken into account.
Moreover, for small values of $\sigma$ (ie $\sigma<5.5$ Mpc/h in our analysis) 
the lensing effect at large $\pi$ separations is dominated by 
the correlation at $r \simeq \sigma$, where the clustering is fully non-linear.
We need to account for these non-linearities if we want to take the full
lensing contribution into account.

We have carried out an investigation of these effects using numerical simulations.
The details are described in Appendix \ref{app:nonlinear}. 
The main conclusions are: {\bf 1.} Eq. (\ref{eq:xigmu})
seems a good approximation to the true magnification bias correction
to the two point function, even after allowing for nonlinear redshift space
effects and their coupling with lensing, provided that {\bf 2.} 
the fully nonlinear
$P_{gm}$ is used in the integral in Eq. (\ref{eq:xigmu}) i.e. nonlinear galaxy bias
is important.
The second point is crucial and bears repeating.
Suppose one is correlating two galaxies that are far apart in the radial direction $\pi$
but close in the transverse direction $\sigma$. 
One might naively think that since the two galaxies are far apart, only linear
galaxy bias needs to be considered.
This is not true once the lensing effect is included.
A small transverse separation means the lensing impact parameter is small.
One is then sensitive to the galaxy-mass correlation on small scales.
Recall that the lensing magnification is an integral of mass overdensity along
the line of sight. The relevant lensing correction $\xi_{g\mu}$ comes from correlating the
galaxy in the foreground with the lensing magnification of the background.
The dominant contribution comes from the portion of the line of sight integral
that is in fact close to the foreground galaxy.
In other words, a theoretical prediction
for $\xi_{g\mu}$ using a linear galaxy bias \citep[such as by][]{hui1} underestimates
the lensing effect for small $\sigma$'s, since the true galaxy bias on small scales
is expected to rise above the linear value. This is illustrated in Fig. \ref{fig:biasnl},
from the analysis in Paper II \citep{paper2}.
The net effect on $\xi_{g\mu}$ is described in Appendix \ref{app:nonlinear},
Fig. \ref{fig:xigmu}. 

When modeling data, we can use Eq. (\ref{eq:xigmu}) with
a nonlinear galaxy bias taken from Fig. \ref{fig:biasnl}. Note that this galaxy
bias, while inferred from data, is strictly speaking not the relevant bias
to use - what we need is the nonlinear bias relevant for galaxy-mass correlation rather
than for galaxy-galaxy correlation. In our modeling in \ref{sec:BAOmodel},
we will therefore allow an extra overall normalization factor multiplying 
Eq. (\ref{eq:xigmu}), which we refer to as $A$.
We will show that the data favors a non-zero magnification bias correction 
$A \neq 0$ at the $2\, \sigma$ level, but our estimate for magnification
bias  $A=1$ is 1.5-sigma smaller from the actual best fit for $A$.
Given the mentioned uncertainties, it is possible the real lensing effect is 
in fact larger than our prediction. 

\begin{figure}
\centering{\epsfysize=5.cm\epsfbox{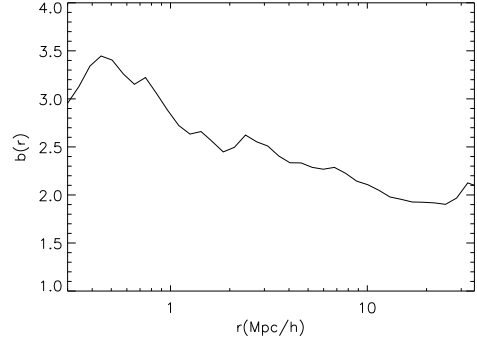}}
\caption{
This is the observed LRG clustering bias as a function of scale $r$, defined by
$b(r) = \sqrt{\xi_{\rm obs} (r)/\xi_{\rm DM} (r)}$, where $\xi_{\rm obs}$ is
the galaxy correlation function and $\xi_{\rm DM}$ is the theoretical
nonlinear dark matter correlation, both in real space.}
\label{fig:biasnl}
\end{figure}

\section{Analysis and Results}
\label{sec:analysis}

\subsection{Data sample}

In this work we use the recent spectroscopic 
SDSS data releases, DR7 and DR6 \citep{dr6}.
We use the same samples and methodology here as presented in Paper I \citep{paper1}
of this series. 
LRG's are targeted in the photometric catalog, via cuts in the (g-r, r-i, r) 
color-color-magnitude cube. Note that all colors are measured using model 
magnitudes, and all quantities are corrected for Galactic extinction 
following \cite{ext1998}.
The galaxy model colors are rotated first to a basis that is aligned with 
the galaxy locus in the (g-r, r-i) plane according to:
$c_{\perp}= (r-i) - (g-r)/4 - 0.18$ and
$c_{||} = 0.7(g-r) + 1.2[(r-i) - 0.18]$.
Because the 4000 Angstrom break moves from the g band to the r band 
at a redshift z $\simeq$ 0.4, two separate sets of selection 
criteria are needed to target LRGs below and above that redshift.
The two cuts together 
result in about 12 LRG targets per $deg^2$ that are not already 
in the main galaxy sample.
The radial distribution and magnitude-redshift diagrams for these
galaxies are shown in Fig.A1 and A12 of Paper I.

We k-correct the r magnitude using the Blanton program 'kcorrect'
\footnote{http://cosmo.nyu.edu/blanton/kcorrect/kcorrect\_help.html}. We need to
k-correct the magnitudes in order to obtain the absolute magnitudes and
eliminate the brightest and dimmest galaxies. We have seen that the previous
cuts limit the intrinsic luminosity to a range $-23.2<M_r<-21.2$, and we only
eliminate from the catalog a small number of galaxies that lay out of the limits. Once we
have eliminated these extreme galaxies, we still do not have a volume limited
sample at high redshift. For the 2-point function analysis we  account for 
this using a random catalog with identical selection function but 20 times
denser (to avoid shot-noise). The same is done in simulations.

There are about $75,000$ LRG galaxies with spectroscopic redshifts 
in the range $z=0.15-0.47$ over $13\%$ of the sky. We break the full
sample into 3 independent subsamples with similar number of galaxies:
low $z=0.15-0.30$, middle $z=0.30-0.40$ and high $z=0.40-0.47$.
The middle sample has a lower amplitude of clustering
because it includes galaxies with a lower luminosity (see Fig.A12 in Paper I).
Also note that the radial distribution is quite peaked in the middle sample (see Fig.A1 in Paper I).
The resulting clustering in this sample has a low signal-to-noise
and we cannot detect the BAO peak in the monopole
(see Fig.31 and A13 of Paper I). We will therefore concentrate on results from
the full sample, as well as the low and high redshift slices for the BAO detection. 

\subsection{2-point correlation}
\label{2ptcorrelation}

The two-point correlation function is defined as
\be{xi2xi3}
\xisp = \langle \delta(r_1) \delta(r_2) \rangle 
\ee
where $\delta(r) = \rho(r)/\bar{\rho}-1$ is the local density fluctuation about 
the mean $\bar{\rho}=\langle\rho\rangle$, and the expectation value is taken over different
realizations of the model or physical process. As mentioned above,
the observed correlation function is anisotropic.
We estimate the correlation as $\xisp$, 
where $\pi$ is the separation along the
line-of-sight (LOS) and $\sigma$ is the transverse separation.
The absolute separation (in redshift space) is
$r_{12}=\sqrt{\sigma^2+\pi^2}$.



In practice, the expectation
value above is over different spatial regions in our universe, which are
assumed to be a fair sample of possible realizations (see Peebles 1980).
A possible complication with this approach is the so-called finite volume
effect which leads to an estimation bias, sometimes referred to as the
integral constraint bias (e.g. see \cite{HuiGazta,bernar}).
For our samples and scale of interest (ie $<150$ Mpc/h) we have checked using a large simulation
that such a potential estimation bias is small compared to the errors.
We use a simulation about 500 times the volume of
the SDSS LRG data (MICEL7680 with 453 ${\,\rm Gpc^3/h^3}$,
$2048^3$ dark matter particles and 107 million halos in a single box; 
see \cite{mice,mice2}) to estimate the ``true'' correlation.
We then split this large simulation into 216 mocks and estimate
the mean and error of the 2 and 3-point correlation 
functions in the 216 subsamples (each similar to the SDSS LRG samples). 
We find that the mean agrees well with the
true value estimated from the full simulation, well within the error estimated from the
dispersion among the subsamples.
In other words, any possible estimation bias, 
on our scales of interest, is much smaller than the errorbar.
The ratio defined by the difference between the true correlation 
(from the full MICE7680 simulation) 
and the mean correlation in 216 smaller 
mocks (cut out of the full MICE7680 simulation)
divided by the dispersion (the error)  indicates how large is the integral constrain bias as compare
to the error. It turns out to be insignificant, always smaller than 0.05\% of the error in both dark matter and halo mocks.
For the monopole we also find that the
integral constraint bias is always smaller than $0.1\%$ of the
error for all the
scales in our analysis, ie $s<150$ Mpc/h (see also paper I and III where we show
several examples of this).

\subsection{Estimation of $\xisp$ and diagonal errors}
\label{estimateError}

To estimate the correlation $\xisp$, 
we use the estimator of \citet{landyszalay},
\begin{equation} 
\xisp  = \frac{DD - 2DR + RR}{RR} 
\end{equation}
with a random catalog $N_R=20$ times denser than the SDSS catalog. 
We find similar results when we use other similar estimators, such as $DD/RR-1$ and $DD/DR-1$ or
estimators based on pixel density fluctuations (ie \cite{barrigagazta2002}). 
The random catalog has the same redshift (radial) distribution as the data, 
but smoothed with a bin $dz=0.01$ to avoid possible cancellation of
intrinsic correlations in the data. The random catalog also has the same mask. 
(We will test robustness against variations in the choice of mask and the radial
selection function below.)
We count the pairs
in bins of separation along the line-of-sight (LOS), $\pi$,  and across the sky,
$\sigma$. The LOS distance $\pi$ is just the difference between the radial comoving distances
in the pair. The transverse distance $\sigma$ is given by
$\sqrt{s^2-\pi^2}$, where $s$ is the net distance between the pair. We use the small-angle
approximation, as if we had the catalog at an infinite distance, which is accurate
until the angle that separates the galaxy pair in the sky 
is larger than about 10 degrees for 
$\xisp$ (see \citet{szapudiwide} and \citet{matsuwide}). This
condition corresponds to transverse scales larger than $\sigma=80$ Mpc/h ($\sigma=165$ Mpc/h) 
for galaxies at z=0.15 (z=0.34).

\begin{figure}
\centering{\epsfysize=6.1cm\epsfbox{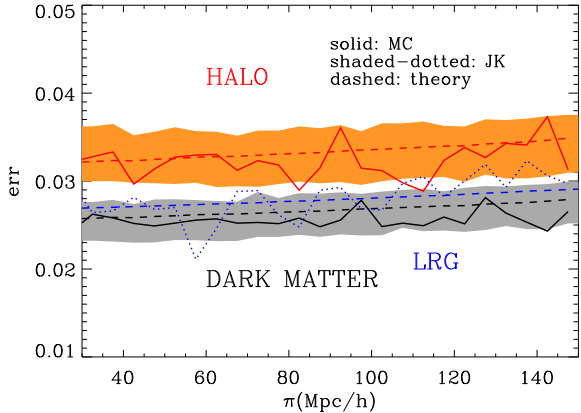}}
\caption{In this figure we show different error estimates for
the 2-point correlation in the 
line-of-sight direction $\pi$, averaged over $\sigma=0 - 5$ Mpc/h.
The black solid line and red solid line shows the dispersion for
dark matter and halos respectively computed
from 216 mock catalogs (the dark matter mocks are at z=0.3,
and the halo mocks are at z=0 chosen with a large scale bias of 1.9).
We also calculate the Jack-knife (JK) error for each mock and we plot 
its dispersion as a shaded region (gray for dark matter and orange for
halos). For comparison, we also show the JK error for the real sample of LRGs
(blue dotted line).
The dashed lines show the error estimates from our analytic model 
for dark matter (black), halos (red) and LRGs (blue) respectively
(see text).
\label{fig:errorvalid2}
}
\end{figure}

There are two sources of error or variance in the estimation of
the two-point correlation: a) {\it shot-noise variance}
which is inversely proportional to the number of
pairs in each separation bin b) {\it sampling variance} which
scales with the amplitude squared of the correlation.
It is easy to check that for the size and number density of our
sample, the shot-noise term dominates over the sampling variance
error. DM particles have a much higher density than LRG galaxies and
for them shot-noise is negligible. We can then dilute the DM particles
to check how and when shot-noise dominates over sampling variance. 
We confirm this using the same simulation MICE7680 mentioned above.
We create out of the large box 216 independent mock catalogs 
with same density than the observed LRG galaxies.
Fig. \ref{fig:errorvalid2} shows the error (square root of the
variance) of the line-of-sight $\xi$ for dark matter (black solid line) and halos (red solid line),
computed from the dispersion between the 216 mocks - this is by definition
the true error. Note that we dilute DM particles to match the observed LRG density
(and N(z) distribution) so now dark matter mocks are also dominated by shot-noise error due to the low density.
We compare this error against two approximations:
Jack-knife error (Gray and orange shaded regions for dark matter and halos
respectively), and an analytic error-estimate (black dashed line and
red dashed line for dark matter and halos respectively). 
The analytic estimate takes the simple form $\Delta\xi= \alpha_{\rm noise}
~\Delta\xi_{\rm Poisson}$, where $\Delta \xi_{\rm Poisson}$ is simply
one over root $N$ shot noise where $N$ is the number of pairs at
the separation bin of interest. For dark matter mocks, $\alpha_{\rm noise} = 1$
works very well, but as discussed in Paper I, the halos or groups do not
quite follow a Poisson distribution and their relevant $\alpha_{\rm noise} = 1.4$. 
The latter is consistent with the findings of others \cite{SmithScocci} that
massive halos, by virtue of exclusion zones around them, do not have Poissonian
shot noise. Fig. \ref{fig:errorvalid2} shows that both Jack-knife and
analytic error estimates work fairly well. For comparison, we also show
the Jack-knife error for the LRGs in real data (blue dotted line) and
the corresponding analytic estimate (blue dashed line).
The agreement between them confirms the validity of our analytic
error model, which we will adopt for the rest of this paper.
Note that the LRG errors are in-between the DM and halo errors because of the
small differences in number density and value of $\alpha_{\rm noise}$.


\subsection{The covariance matrix}
\label{covariance}

\begin{figure}
\centering{\epsfysize=6.cm\epsfbox{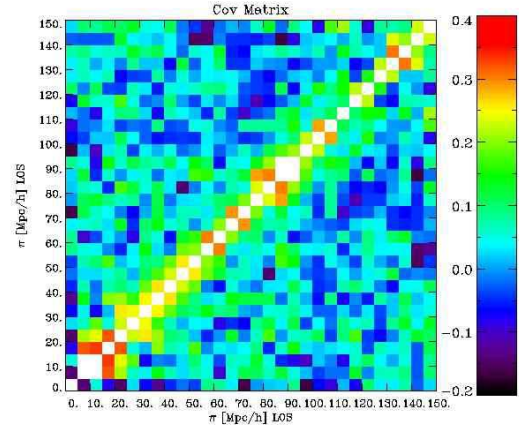}}
\caption{Covariance matrix (relative to the variance) for
LOS correlation in 5 Mpc/h bins. Even if the covariance is
not negligible close to the diagonal, results in our analysis
are very similar when we assume a diagonal matrix.
\label{fig:covar}
}
\end{figure}

We also use the galaxy (halo) mocks to estimate the covariance matrix between bins
in the whole $\xips$ plane. Figure \ref{fig:covar} shows the covariance
for 5 Mpc/h bins in the LOS, normalized to the corresponding 
variance (ie diagonal elements are unity). 
Full details of this estimation are given in Paper I.
The figure shows that at scales $\pi>20 Mpc/h$ the covariance is quite
small. It is smaller than 10\% in most of the plane, except in the bins
closer to the diagonal where it can be larger but still lower 30\%. 
At the BAO position  $\pi \simeq 110$ Mpc/h the covariance is lower than 
10\% for all bins. We do  our analysis using this covariance matrix,
but we have checked that results are very similar when we assume a diagonal
matrix (eg see Fig.\ref{fig:baos2nall}), so that in practice the covariance 
can be neglected.

\subsection{Results for $\xisp$}

\begin{figure*}
\centering{\epsfysize=5.cm\epsfbox{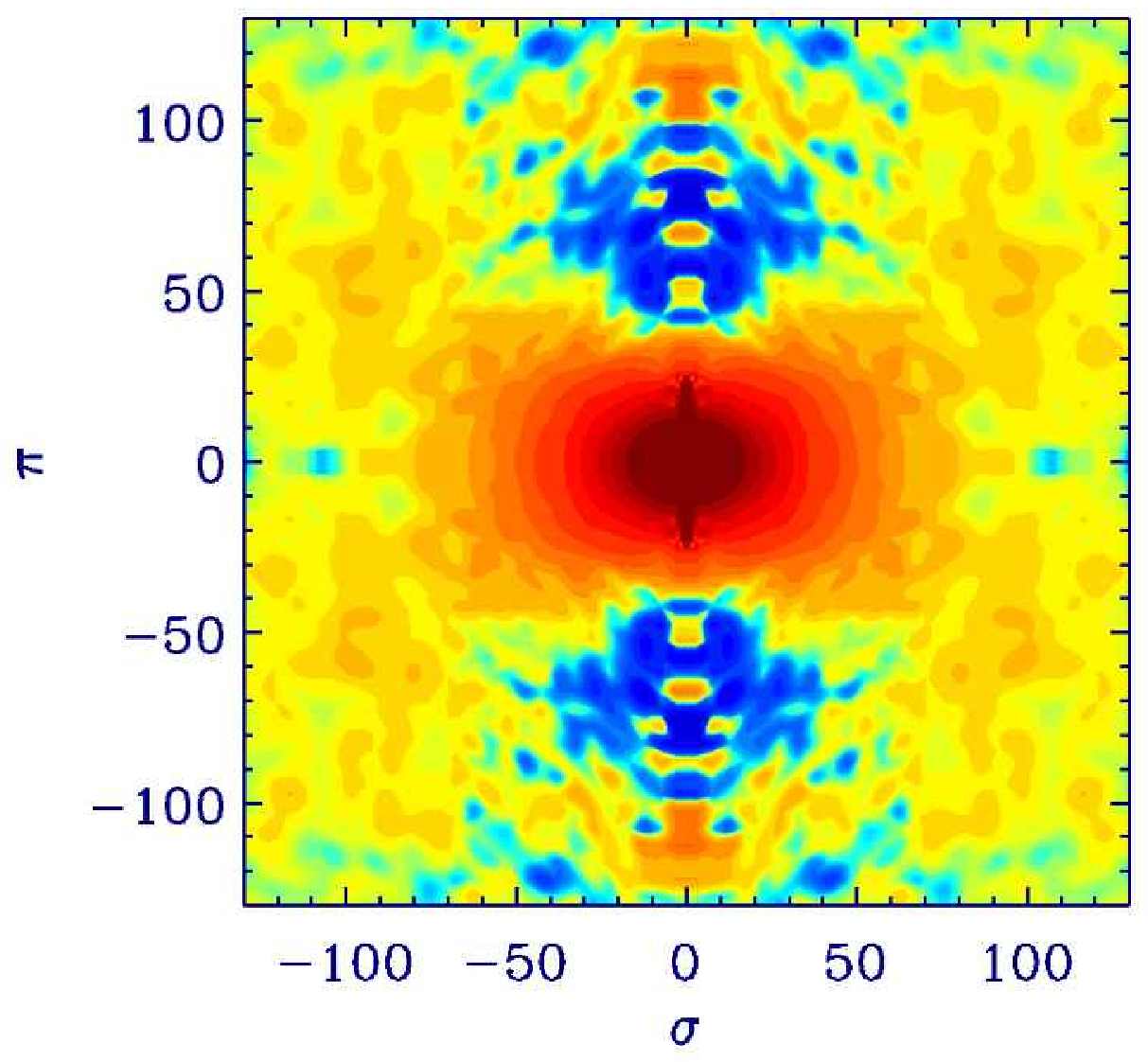}}
\centering{\epsfysize=5.cm\epsfbox{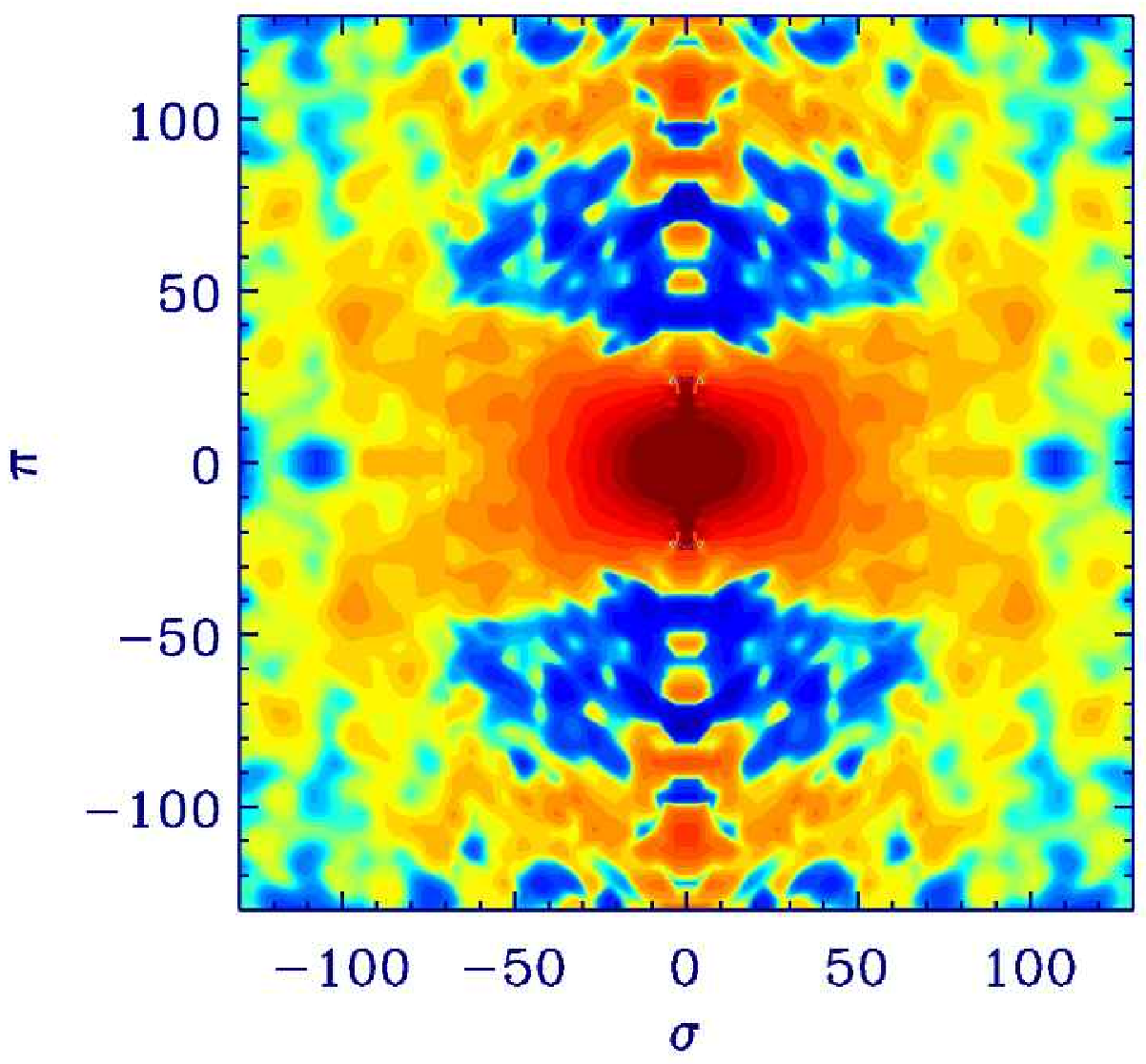}}
\centering{\epsfysize=5.cm\epsfbox{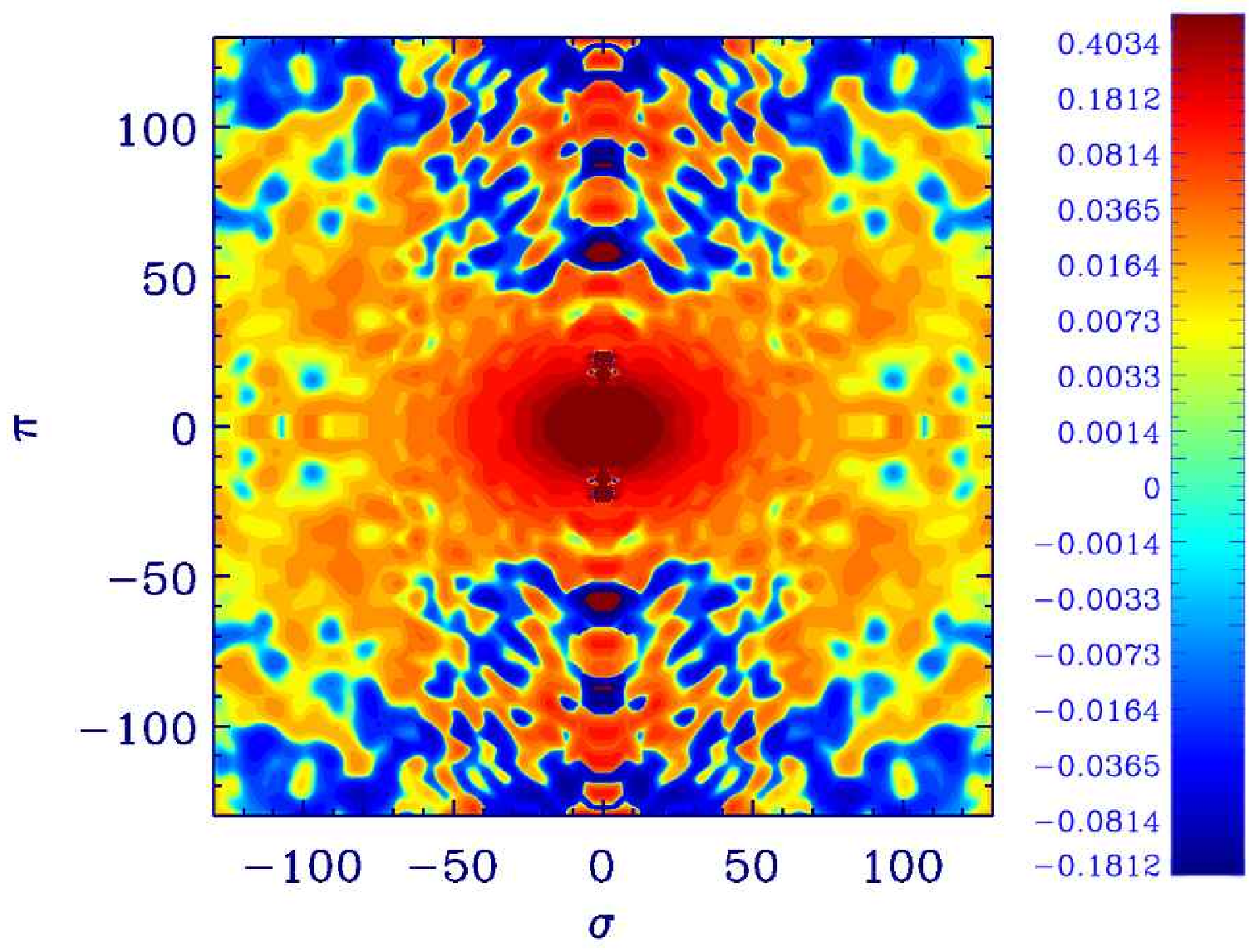}}
\caption{Measurements of $\xips$ from different redshift slices. 
From left to right: z=0.15-0.47 (all), z=0.15-0.30 and z=0.40-0.47.
\label{fig:magnis2} }
\end{figure*}

In Fig.\ref{fig:magnis2}
we see the redshift-space correlation function $\xisp$ for the complete 
catalog (z=0.15-0.47), and for two different slices in redshift: 
$z=0.15-0.3$ and $z=0.40-0.47$.
Recall from the top panel of Fig. \ref{fig:baoteoric1} that,
without magnification bias, the conventional expectation is that
one should see a less prominent BAO peak in the LOS direction $\pi$ than
in other directions. Instead, we see from the data, Fig. \ref{fig:magnis2},
that the observed BAO peak actually gets more pronounced
along the LOS direction, in qualitative agreement with the bottom panel
of Fig. \ref{fig:baoteoric1} which includes magnification bias.
Indeed, we see a very nice ring associated with the BAO peak in the data.
Note also in the data, we see these valleys of negative correlations (blue) on scales of
$\pi = 50-90$ Mpc/h, which are in accord with the predictions
of Kaiser distortions (Fig. \ref{fig:baoteoric1}).

In separating $\sigma$ from $\pi$ in the data we have assumed the plane-parallel approximation. 
This introduces a distortion of the BAO scale in the 
perpendicular direction $\sigma > 100$ Mpc/h
when $\pi$ is small. This can be clearly seen in the plots, where there is an artificial
concentration of the signal at an angle of a few tens of degrees away from the $\pi=0$ axis,
which produces an X shape in our $\sigma-\pi$ diagrams, especially at large $\sigma$'s.
In reality, this signal originates from smaller angles.
This effect is explicitly demonstrated in our simulations
\citep[see Fig.A14 of Paper I][]{paper1}.
In the Appendix of Paper I  we show how to correct for this effect by removing some of the
pairs in the calculation. This is not the best possible
approach, since it throws away information, but it shows
the origin of this strange X shape feature at large $\sigma$'s.
\citep[see also Fig.9 in]{matsuwide}. 
Note, however, that neither 
the LOS clustering ($\sigma = 0$) nor the monopole, studied here,
are affected by this artificial distortion.

We use a fiducial flat model with matter density $\Omega_m=0.25$ 
and Hubble constant $h=0.72$ to convert the observed redshifts and angles into distances.
Unless stated otherwise we also assume baryon density $\Omega_b=0.045$, 
and spectral index $n_s=0.98$ as
the concordance LCDM model ($w=-1$ for the dark energy equation of state).

\subsection{The Monopole}
\label{sec:monopole}

\begin{figure*}
\centering{\epsfysize=4.0cm\epsfbox{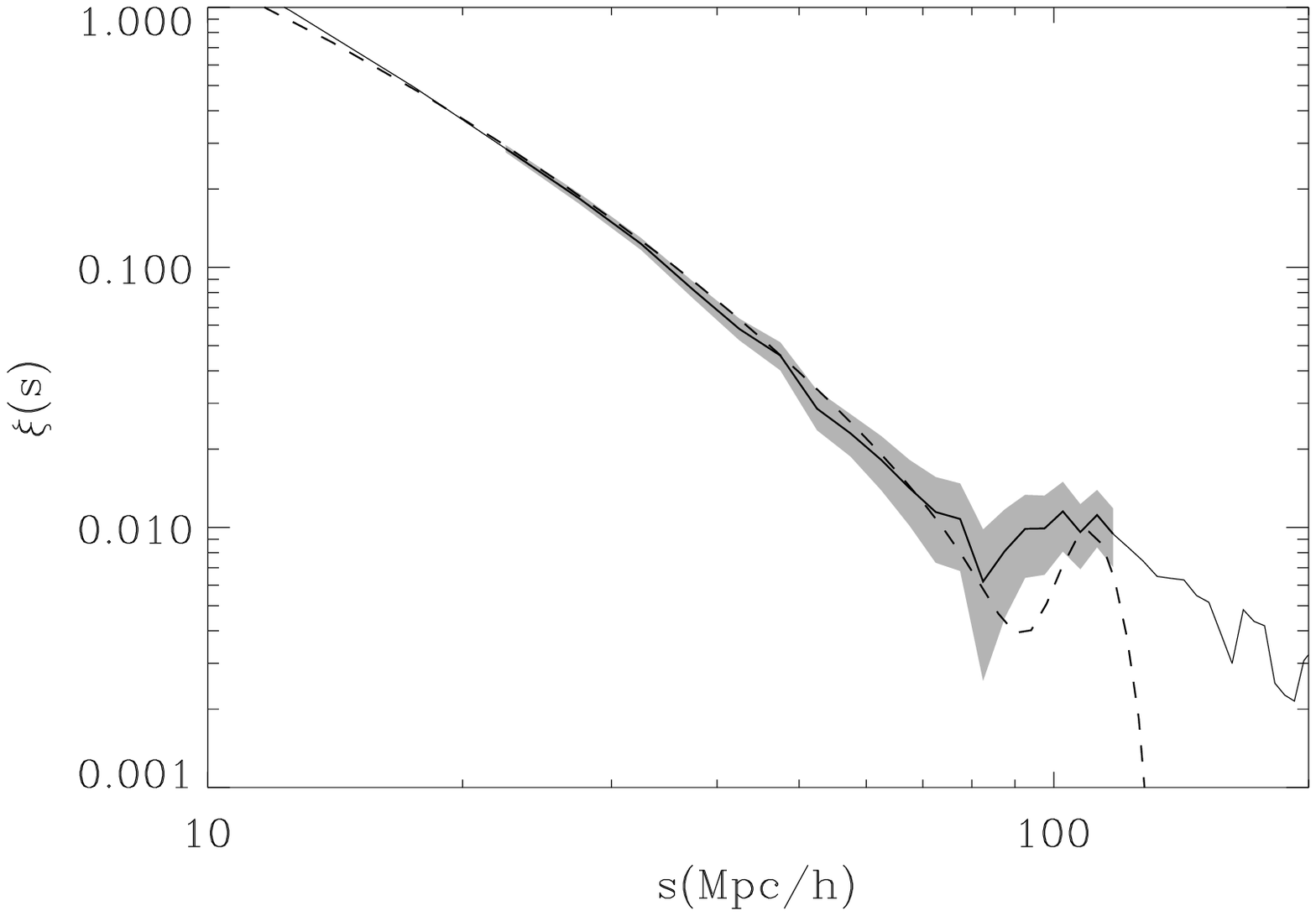}}
\centering{\epsfysize=4.0cm\epsfbox{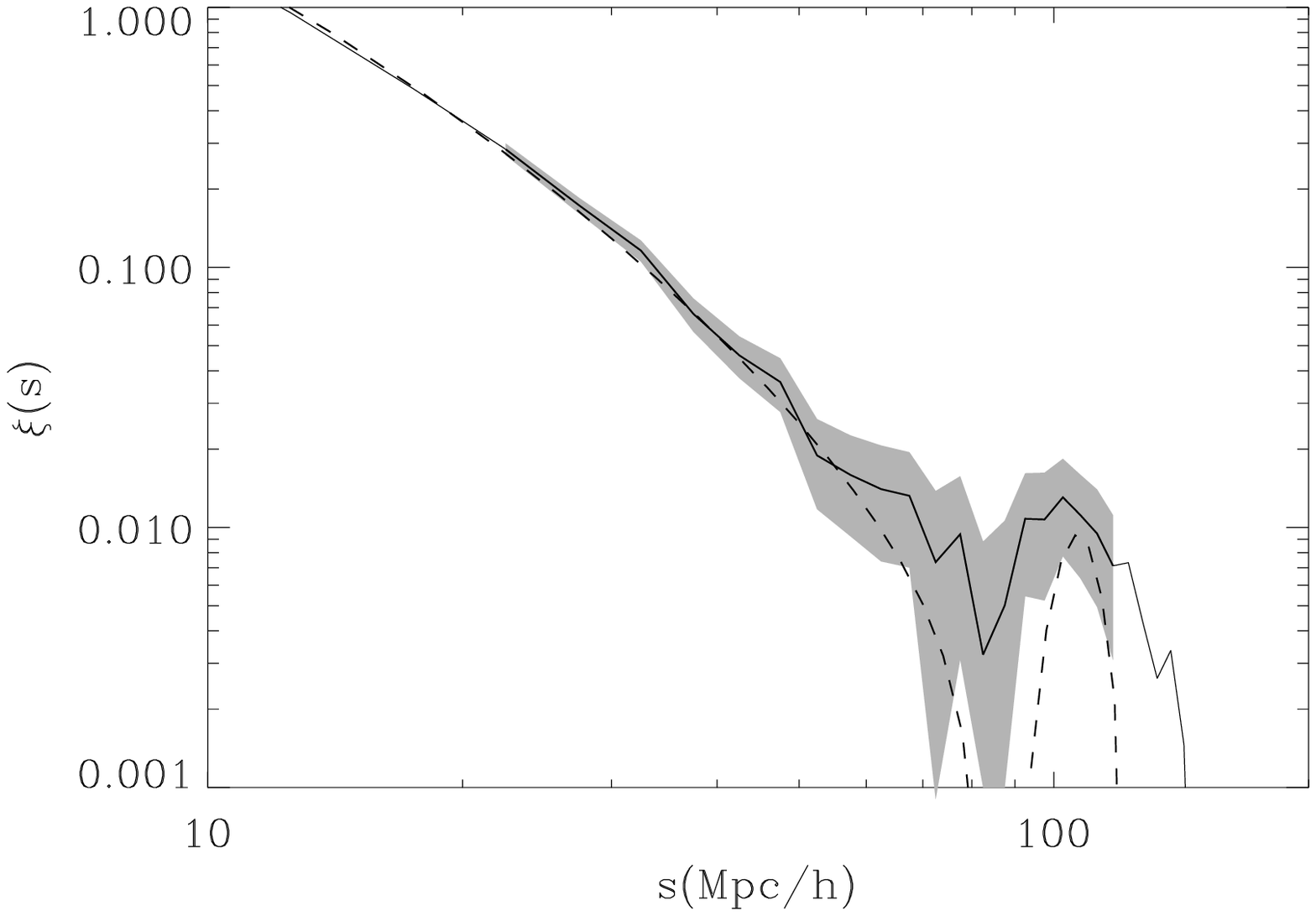}}
\centering{\epsfysize=4.0cm\epsfbox{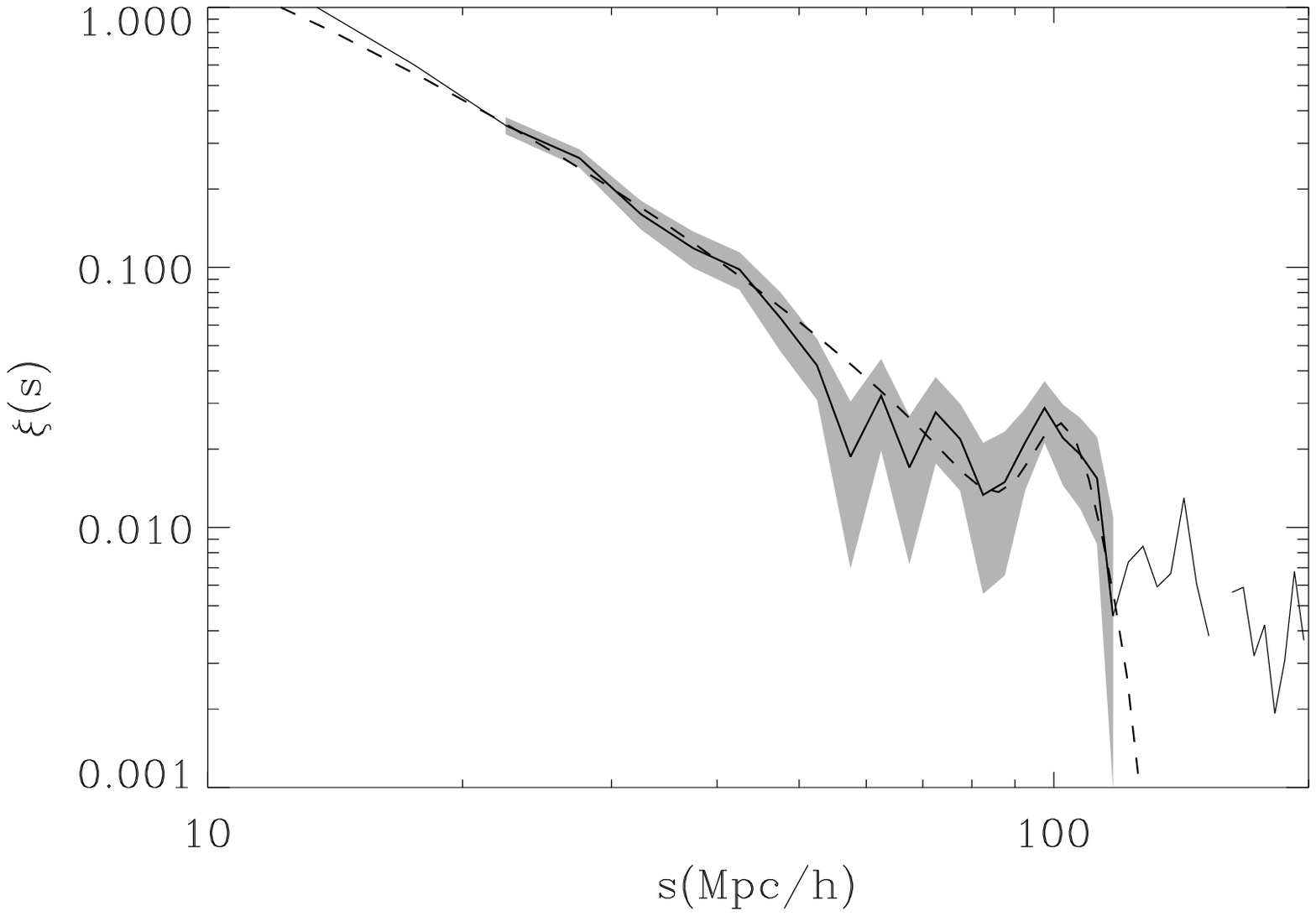}}
  \caption{Measured monopole with errors (solid line with
gray area) compared with the best fit model (dashed line), which
uses the scales $20 - 120$ Mpc/h (20 log bins).
 {\it Left panel:} corresponds to the full sample (z=0.15-0.47),
where we 
find $\chi^2=3.4$ (best
fit is $\Omega_m h^2=0.12$, $\Omega_b h^2=0.026$).
{\it Middle panel:} corresponds to the slice
 z=0.15-0.30, where we 
find $\chi^2=1.9$ (best
fit is $\Omega_m h^2=0.132$, $\Omega_b h^2=0.028$).
{\it Right panel:} corresponds to the slice z=0.40-0.47
with $\chi^2=4.8$
(best fit is $\Omega_m h^2=0.124$, $\Omega_b h^2=0.04$).
Allowed values are shown in Fig.\ref{fig:fit.slice}.
 \label{fig:xs.slice}}
\end{figure*}

\begin{figure*}
\centering{ \epsfysize=4.35cm\epsfbox{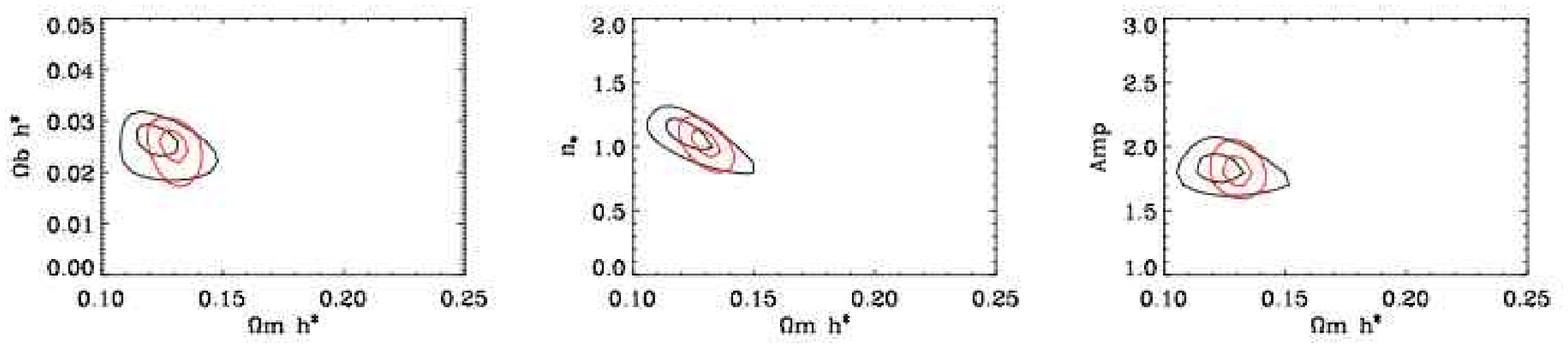}}
\centering{ \epsfysize=4.35cm\epsfbox{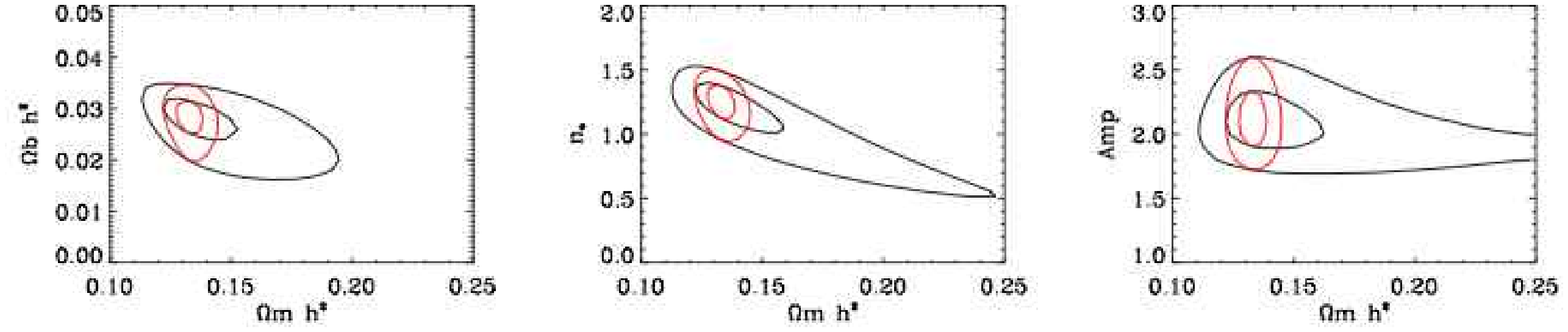}}
\centering{ \epsfysize=4.25cm\epsfbox{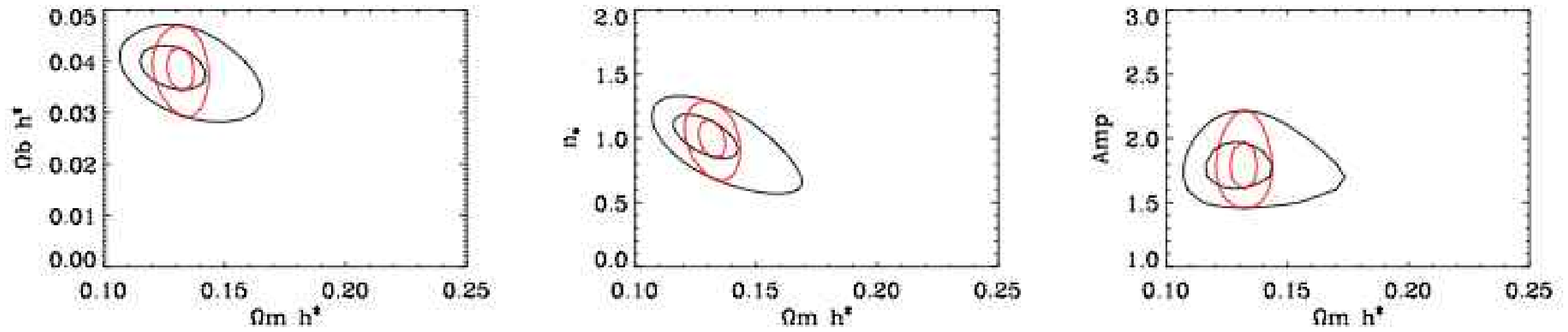}}
\caption{Best fit contours (1 and 2-sigma with 1dof) for cosmological parameters
in a fit to the monopole. Top, middle and bottom panels correspond to redshift slices z=0.15-0.47,
z=0.15-0.30 and z=0.40-0.47 respectively. The smaller inner (red) contours
use a prior of $\Omega_m h^2= 0.1326 \pm 0.0063$ from WMAP5.
 \label{fig:fit.slice}}
\end{figure*}

As a first step in studying the BAO, we look at the monopole, which is
the average of $\xi(\sigma, \pi)$ over orientations:
\begin{eqnarray}
\label{xi0}
\xi_0 (r) = \int_{-1}^1 \xi(\sigma, \pi) {d\mu\over 2}
\end{eqnarray}
where $r = \sqrt{\sigma^2 + \pi^2}$ and $\mu = \pi/r$.
Fig. \ref{fig:xs.slice} shows the observed monopole (solid line with
gray area denoting errorbar) in different redshift slices. 
The BAO peak is clearly visible. One way to verify its significance
is to perform a parametric fit and see if a non-zero baryon density
$\Omega_b$ is required by data.
We use 4 parameters: $\Omega_m h^2$, $\Omega_b h^2$, $n_s$ and an overall
amplitude $Amp$.
Our model is essentially linear theory, but includes
non-linear effects according to re-normalized perturbation theory (RPT) at the BAO peak, 
as described in \citet{croccebao}. Magnification bias is negligible for the monopole.
As is demonstrated in Fig. 10 of Paper I, the ratio of the monopole to
the real space correlation function is constant on scales larger than
about $10$ Mpc/h, consistent with the Kaiser model. This is why
we fit the data with a single amplitude $Amp$ which is supposed to
account for $\sigma_8$, galaxy bias and the redshift distortion boost,
all rolled into one. To be conservative, only data on scales larger than
$20$ Mpc/h are used in our fit. Covariance between different scales
is taken into account, using the error model developed and tested in Paper I.
Magnification bias is not included in our fits because its effect on the monopole
is quite small at our moderate redshifts (\citet{hui1}).

The resulting constraints are shown in Fig. \ref{fig:fit.slice}, and
the corresponding best-fit monopole is shown as a dashed line in Fig. \ref{fig:xs.slice}.
The $\chi^2$ values quoted in the caption are obtained by applying singular value
decomposition to the covariance matrix, keeping modes with the smallest 7 eigenvalues, though
the fits are robust against varying this number.

In order to calculate the covariance, we have taken 216 mock catalogs with the same 
characteristics than LRG galaxies. Moreover, we can 
calculate the JK covariance for each mock and compute the mean value over 
all the mock catalogs. This covariance is very similar to the Monte Carlo 
one but smoother so we use the JK one to calculate $\chi^2$. 
This is extensively explained in Appendix A3 of Paper I.

\begin{figure*}
\centering{\epsfysize=5.cm\epsfbox{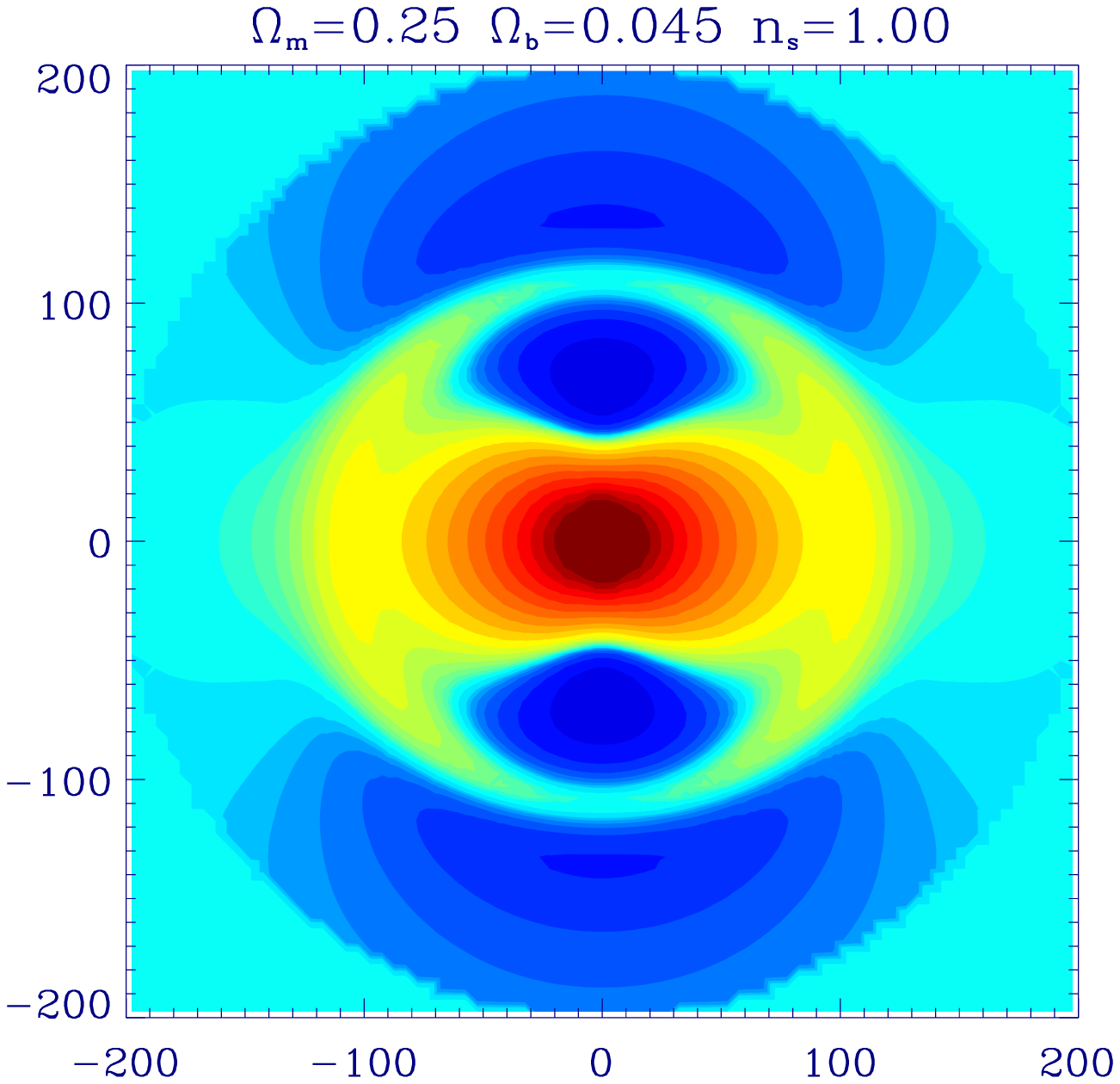}}
\centering{\epsfysize=5.cm\epsfbox{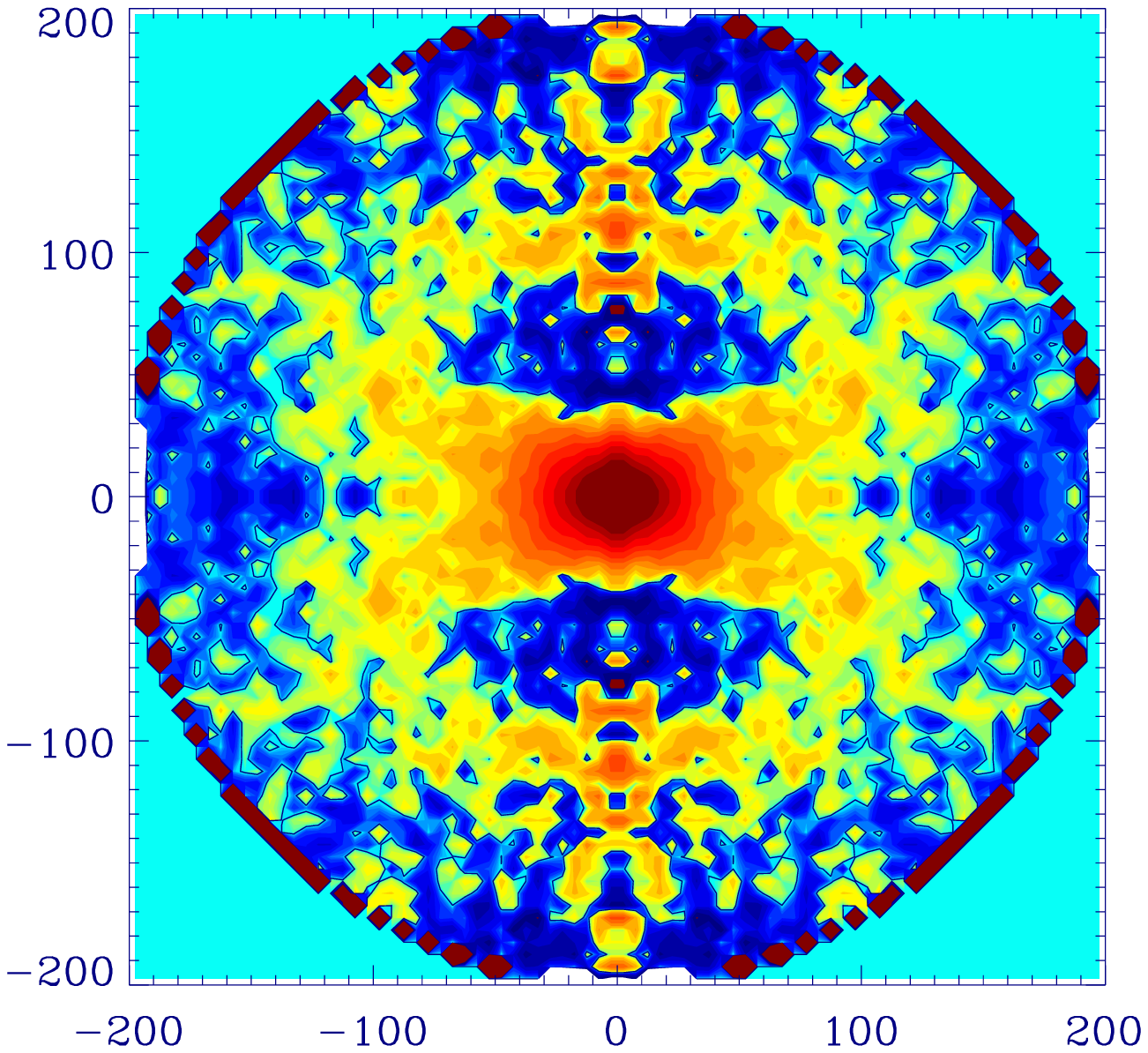}}
\centering{\epsfysize=4.7cm\epsfbox{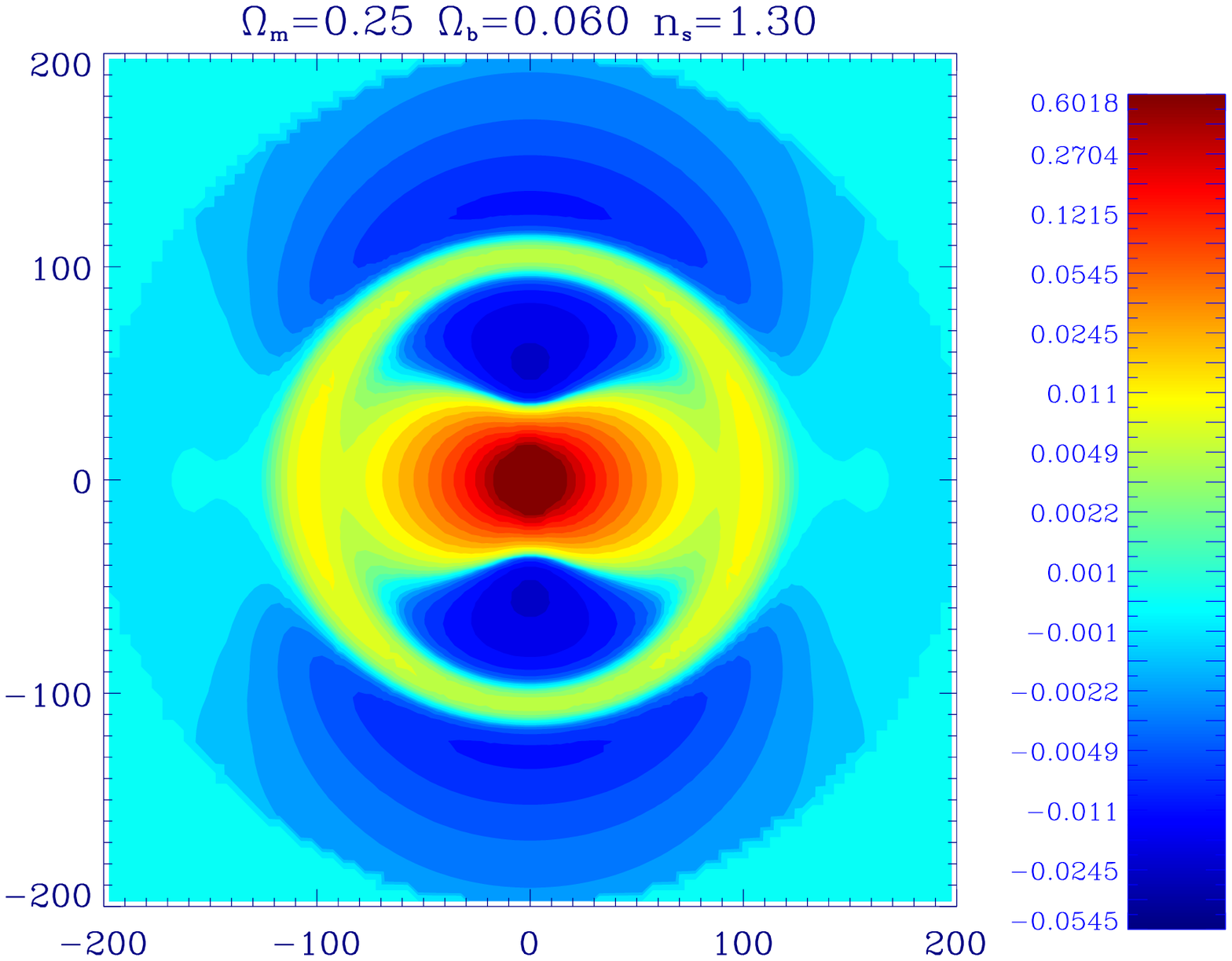}}
\caption{A comparison of $\xisp$ in data and models for the z=0.15-0.30 slice.
Left panel corresponds to a standard cosmology (low $\Omega_b=0.045$ and $n_s=1.0$) model, 
while the right panel has a more prominent BAO peak ($\Omega_b=0.06$ and $n_s=1.3$), which corresponds
to the best fit monopole model to the same data. 
Middle panel shows the data using the same color scheme. 
\label{fig:pisigma}}
\end{figure*}

There is some tension between our best fit $\Omega_b$ (or $\Omega_b h^2$) and 
the WMAP5 value \cite{wmap5}
for the same model (flat universe with  w=-1).
Our best fit values tend to be higher. A higher $\Omega_b$ leads to a more prominent
BAO peak, as illustrated in Fig. \ref{fig:pisigma}. 
However, it should be emphasized that in the low redshift slice $z=0.15 - 0.30$ and
in the full sample $z = 0.15-0.47$ 
the best fit $\Omega_b h^2$ is less than 2-sigma away from the standard WMAP5 value of
$0.22$, though the discrepancy is larger in the high redshift slice $z=0.40 - 0.47$.
It is also worth noting that there is more room for accommodation once more
parameters are allowed to vary, such as the dark energy equation of state $w$
and the neutrino mass. A more detailed analysis of this is presented in \citet{Sanchez09}.

\subsection{The BAO Ring}

\begin{figure}
\centering{\epsfysize=7.cm\epsfbox{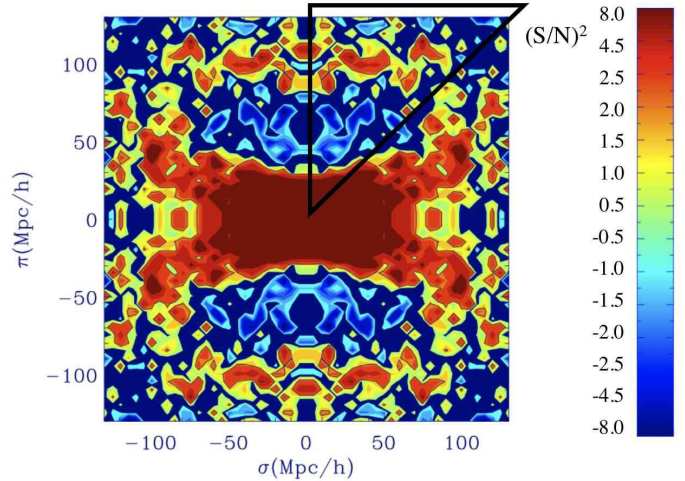}}
\caption{Signal-to-noise ratio in $\xisp$
for the z=0.15-0.30 slice. The color scheme denotes
$(S/N)^2$ multiplied by the sign of the signal i.e.
negative values correspond to a negative signal. The triangle
highlights the region $\pi>\sigma$, which receives little weight in the monopole.
\label{fig:s2nps.11}}
\end{figure}

Recall from Eq. \ref{xi0} that
the monopole receives most of its weights from orientations close to the
transverse direction (i.e. the measure $d\mu$ equals ${\rm \,sin}\theta d\theta$, where
$\theta$ is the angle with respect to the radial direction).
Let us therefore consider briefly the reality of the BAO peak
in other directions, including those close to the LOS or radial direction.
In Fig. \ref{fig:s2nps.11}, we show the signal-to-noise of $\xi$ in the
$\sigma-\pi$ plane for the redshift slice $z=0.15 - 0.3$.
This complements the $\xisp$ signal plot in the middle panel of Fig. \ref{fig:pisigma}.
The signal-to-noise shown in Fig. \ref{fig:s2nps.11} is for
each pixel of size $5$ Mpc/h by $5$ Mpc/h
(the same pixel size is used throughout this paper in all of our
$\xisp$ plots). Note that there is covariance between pixels, and so 
this figure should be interpreted with some care (see Paper I).
Nonetheless, it demonstrates the high quality detection of a BAO ring in
the $\sigma-\pi$ plane. The triangle highlights
the region $\pi>\sigma$, which receives not much weight in the monopole, but
where the BAO ring still shows up nicely.
Note that the $(S/N)^2$ shown is modulated by the sign of the signal:
the (blue) valley of negative correlations at $\pi \sim 50 - 90$ Mpc/h
- in accord with the predictions of the Kaiser effect - are detected
with significance as well. The overall coherent structure of a negative valley
before a positive BAO peak (at just the right expected scales) 
is quite striking, and cannot be easily explained away
by noise or systematic effects.

Fig. \ref{fig:s2nps.11} suggests that there is sufficient information in
the data to separately constrain the angular diameter distance $D_A (z)$ and the
Hubble expansion rate $H(z)$, as discussed in \S \ref{intro}.
That is to say, there is in principle the exciting possibility of determining
both quantities, beyond measuring the combination $(D_A^2 / H)^{1/3}$ from
the monopole as is customarily done (\citeauthor{detection} \citeyear{detection}; 
see also Appendix A of \citeauthor{hui2} \citeyear{hui2}).
In this paper, we will focus on the derivation of $H(z)$ from clustering in the radial
direction, in part because, as discussed earlier in \S \ref{sec:analysis}, the plane parallel
approximation that we have adopted introduces artificial distortions to
the signal at large $\sigma$'s. This does not affect the analysis here which focus on LOS
and monopole but will have an impact in $\xips$.
Modeling the wide-angle effects is possible,
and is something we hope to pursue in the future.

On scales larger than 120 Mpc/h we also find some excess (i.e. away from null)
signal but the amplitude is lower and the sign alternates
between positive and negative values and is less coherent than the
positive or negative regions at smaller scales.
This excess signal on larger scales
is also found in the monopole for some of the redshift slices.
In the Appendix of Paper I  we look for possible systematic effects that
could produce this excess. By changing the mask in extreme ways, it is possible to 
reduce the amplitude of these excess fluctuations on very large scales 
(see Fig.A8 in Paper I), but these variations do not change the location of the BAO peak,
as will be discussed below.

\subsection{The Radial Peak}
\label{radialpeak}

In this section, we turn our focus to the correlation function in the
LOS/radial direction. A visual impression of the LOS $\xi$ in the full sample
can be gained from Fig. \ref{fig:LOSsys}, \ref{fig:losdirectionmodel2} and Table \ref{tab:data}, which
will be explained in more details below. What we would like to do first
is to check for systematics and test the robustness of our measurement.
Then, we will examine the statistical significance of the BAO feature.

\subsubsection{Tests for Systematics}
\label{sec:SystematicsTests}

There are many checks that need to be made, some of which we have already
mentioned. Here, we provide a complete list.

{\bf 1.} Finite volume effects or integral constraint bias, namely
an estimation bias that results from the galaxy survey having a finite size,
is examined in \S \ref{2ptcorrelation} using large numerical simulations.
We find that for the scales of interest, the integral constraint bias
is totally negligible compared to errorbars.

{\bf 2.} We also employ numerical simulations to test the accuracy of
our (statistical) error model. This is done in \S \ref{estimateError}
(see Fig.\ref{fig:errorvalid2}).
We find that our error model agrees well with the Jack-knife error
and also, in the case of simulated data, with the true error estimated from
Monte Carlo realizations.

\begin{figure}
\centering{\epsfysize=7.5cm\epsfbox{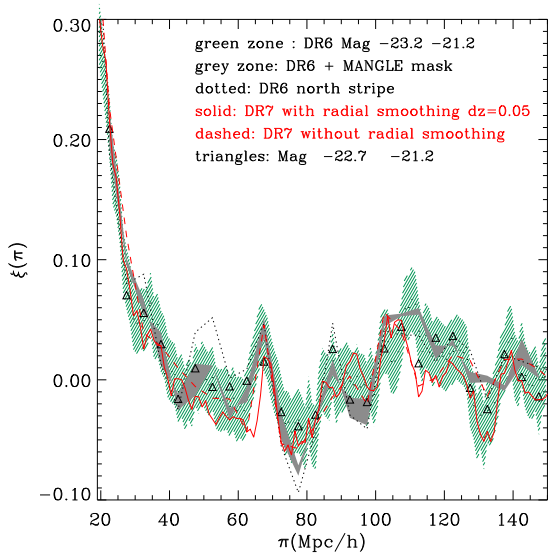}}
\caption{
This figure shows $\xi(\pi)$ measured using different radial selection
functions, masks and datasets, for $z = 0.15 - 0.47$. Details are described in the text.
Here, the LOS correlation is binned in $(\sigma, \pi)$ pixels of $5$ Mpc/h by $5$ Mpc/h.
Each pixel has $\sigma$ extending from $0.5$ Mpc/h to $5.5$ Mpc/h
(the minimum $\sigma$ of $0.5$ Mpc/h is imposed to avoid the fiber collision zone).
The LOS correlation is plotted centered around $\pi$ increments of $1$ Mpc/h - 
there is therefore an overlap between pixels in the $\pi$ direction. This is
purely for presentation purpose, to show how results change for different bin
center positions.
\label{fig:LOSsys}}
\end{figure}

{\bf 3.} We test for robustness against a different choice of the radial
selection function. In Fig. \ref{fig:LOSsys}, the green zone
shows the one sigma region of our fiducial measurement from DR6. 
This fiducial measurement, which is used in the rest of this paper,
uses a radial selection function which is exactly the observed $dN/dz$
from data but smoothed in bins of $dz=0.01$. 
The red solid and dashed lines in the same figure show measurements from
Data Release 7 (DR7), with different choices of the radial selection function
- the solid line uses a radial selection smoothed with
$dz = 0.05$; the dashed line uses a radial selection that is not smoothed
at all i.e. it uses $dN/dz$ straight from the data.
We can see that the resulting $\xi(\pi)$ is, to within errors, stable against
these different assumptions about the actual radial selection function.

{\bf 4.} We test for robustness against a different choice of the angular mask.
After the release of DR6, 
\citet{swanson08} provided mask information in a readily usable form, translating 
the original mask files extracted from the NYU Value-Added Galaxy Catalog \citep{blanton2005}, 
from MANGLE into Healpix format \citep{healpix}. Paper I describes 
how they constructed a survey "mask" for LRGs and tested the impact of the mask on 
clustering measurements using mock catalogs.
Here, we apply the MANGLE mask with different completeness 
in place of our fiducial mask  to DR6, and
the result is shown as the gray zone in Fig. \ref{fig:LOSsys},
which encompass the whole range of possible completeness above zero.
Our measurement appears to be stable against variation in the mask.

{\bf 5.} We test for the stability of our measurement when we look at
different subsets of the data. The triangles in Fig. \ref{fig:LOSsys}
show $\xi(\pi)$ measured from a set of DR6 LRGs with a different magnitude cut.
The dotted line shows $\xi(\pi)$ measured from the north stripe of DR6.
Finally, the red lines show measurements from DR7, which is about 
$\sim 17 \%$ larger than DR6. Again, they are all consistent with each
other to within errors.

In Table \ref{tab:data} we list our best estimates and errors for the measurement of
 $\xips$ along the LOS for our default DR6 mask. This corresponds to the green 
shaded region in Fig.~\ref{fig:LOSsys}. Because errors are correlated on scales
smaller than 5 Mpc/h we give the measurements and errors in the Table smoothed with a top-hat window
of total width 5 Mpc/h, this makes the analysis quite insensitive to the choice of
the central bin position. The resulting covariance between these 5 Mpc/h bins is 
negligible in practice (see Fig.\ref{fig:covar}) and the
different points in the figure can be treated as independent. 
See also Appendix 2.4 in paper I for an extensive treatment of the covariance.

\begin{table}
\begin{center}
 \begin{tabular}{ c r r }
\hline
$\pi$ (Mpc/h) & $\xi(\pi,\sigma=3)$ & error \\
\hline
         2.5 &      5.870 &    0.0253 \\
         7.5 &      2.219 &    0.0254 \\
        12.5 &     0.8517 &    0.0254 \\
        17.5 &     0.3851 &    0.0255 \\
        22.5 &     0.2052 &    0.0256 \\
        27.5 &     0.0883 &    0.0256 \\
        32.5 &     0.0487 &    0.0257 \\
        37.5 &     0.0308 &    0.0257 \\
        42.5 &    -0.0020 &    0.0258 \\ 
        47.5 &    -0.0037 &    0.0259 \\
        52.5 &    -0.0067 &    0.0259 \\
        57.5 &    -0.0183 &    0.0260  \\
        62.5 &    -0.0077 &    0.0261 \\
        67.5 &     0.0245 &    0.0261 \\
        72.5 &    -0.0228 &    0.0262 \\
        77.5 &    -0.0353 &    0.0263 \\
        82.5 &    -0.0277 &    0.0264 \\
        87.5 &    -0.0017 &    0.0265 \\
        92.5 &    -0.0014 &    0.0266 \\
        97.5 &    -0.0042 &    0.0267 \\
       102.5 &     0.0218 &    0.0268 \\
       107.5 &     0.0407 &    0.0268 \\
       112.5 &     0.0454 &    0.0269 \\
       117.5 &     0.0244 &    0.0270 \\
       122.5 &     0.0226 &    0.0271 \\
       127.5 &    -0.0049 &    0.0272 \\
       132.5 &    -0.0274 &    0.0273 \\
       137.5 &     0.0128 &    0.0274 \\
       142.5 &     0.0073 &    0.0275 \\
       147.5 &    -0.0001 &    0.0276 \\

\hline
\end{tabular}
\caption{Our best estimates of the 2-point correlation $\xips$ in the LOS for
$0.5 <\sigma < 5.5$ ( Mpc/h). In the radial direction the correlation is binned
with $\Delta \pi=5$ Mpc/h and we take the average correlation within that bin. 
The covariance between bins is negligible when using this binning. This is for
$z = 0.15 - 0.47$.
\label{tab:data}
}
\end{center}
\end{table}

\subsubsection{Model fitting: shape constraints}
\label{sec:BAOmodel}

\begin{figure}
\centering{ \epsfysize=7.5cm\epsfbox{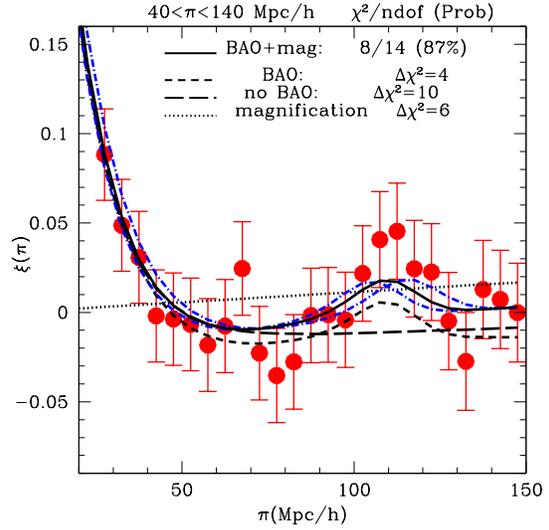}}
\caption{
Points with  1-sigma errorbars show the measured LRG correlation $\xisp$ along 
the LOS (for $0.5 < \sigma < 5.5$ Mpc/h) 
in the full SDSS sample (z=0.15-0.47) as given in Table \ref{tab:data}.
The solid black line is the best-fit model as described in text.
The dotted black (straight) line shows the contribution from magnification bias.
The dashed black line with no-BAO feature is a no-wiggle model.
The dashed black line with a BAO feature is a no-magnification-bias model.
The $\chi^2$ for the best-fit, and the $\Delta\chi^2$ for the two dashed lines
are given at the top of the figure. The dot-dashed blue lines show
the $1-\sigma$ range allowed by a shift $D_r$ in the radial scale $\pi$.
\label{fig:losdirectionmodel2}}
\end{figure}

We next want to check how well the data agree with the theoretical model
presented in previous sections. In particular, we are interested in assessing the
statistical significance of any BAO detection. 
This is illustrated in 
Fig.\ref{fig:losdirectionmodel2}. The observed correlation function (points
with errorbars) is shown in
radial bins of $\Delta \pi=5 {\,\rm Mpc/h}$ and $0.5 <\sigma < 5.5$ Mpc/h.
We fit the data with a model consisting of the following ingredients:
the real space power spectrum is controlled by
$\Omega_B$ and $\Omega_m$ with an amplitude parametrized by $b \sigma_8$,
where $b$ is the linear galaxy bias; redshift space distortion is described
by a Kaiser model parametrized by $\beta$ (nonlinear pairwise velocity
dispersion is kept fixed at $400$ km/s but its effects are important only for scales
less than $\pi < 40$ Mpc/h); the magnification bias correction is linear in 
$\pi$ (Eq. [\ref{eq:xigmu}]), parametrized
by an overall normalization $A$ (as described in \S \ref{sec:nonstandard}).
Note that we $A=1$ corresponds to $s=1.5$. As we allowed $A$ to varied this
corresponds to degenerate changes in $s$ or other normalization factors.
The parameters $\Omega_B$, $\Omega_m$, $b \sigma_8$
and $\beta$ are constrained by the observed monopole and quadrupole of 
$\xi(\sigma,\pi)$  (Paper I).
We therefore impose the following ($1-\sigma$ Gaussian) priors: 
$\Omega_B = 0.044 \pm 0.003$, $\Omega_m = 0.245 \pm 0.020$, $b \sigma_8 = 1.56 \pm 0.09$
and $\beta = 0.34 \pm 0.03$ (the prior on $\Omega_B$ is from WMAP5, while the rest is
from Paper I). We hold fixed 
$h = 0.72$, $\sigma_8 = 0.85$ and $n_s = 0.96$, which are mostly
degenerate with the other parameters, but verify that
relaxing them in a manner consistent with current data 
does not alter our conclusions significantly.
We allow $A$ to vary anywhere between $-1$ to $5$ (flat prior).
Lastly, we also 
allow for a shift $D_r$ in the radial scale 
$\pi$ to account for the fact that
we do not know the true value of $E(z) \equiv H(z)/H_0$ to convert the measured redshift
into the radial distance $\pi$. Note that we can take this to be independent of 
$\Omega_m$ in $H(z)$, even for a flat universe, 
as this would correspond to variations in the DE equation of state $w$ which does not
alter the shape of the correlation on linear scales. We define 
\begin{equation}
D_r\equiv {H_f (z) \over{H(z)}} ={E_f(z) \over{E(z)}} =
{{\sqrt{0.25(1+z)+0.75}} \over E(z)} 
\label{eq:shift}
\end{equation}
where $E_f(z)$ corresponds to the fiducial flat LCDM cosmology with $\Omega_m=0.25$
used in our analysis. 

The best fit model (to the full sample) 
is shown as a solid black curve in Fig. \ref{fig:losdirectionmodel2}.
Here, the range of scales on which the fit is performed is
$\pi = 40 - 140$ Mpc/h.
The best $\chi^2$ of the fit is $\chi^2 \simeq 8$ for 14 
degree of freedom (as labeled in the figure), 
which is reasonable (gives a probability of $Prob=87\%$).
We can use this model fitting procedure to address the reality
of the BAO feature: is it supported by data? 
A model with no BAO  \citep[generated according to][]{eisensteinhu} 
leads to a $\Delta \chi^2 = 10$ away
from the best-fit i.e. it is ruled out at the
$3.2\, \sigma$ level.

It is also interesting to ask 
to what extent the magnification bias correction is required.
We find the magnification bias normalization $A = 4.5 \pm 2.2$. 
Another way to put it is that
a model with no magnification bias leads to a
$\Delta \chi^2 = 4$ away from the best-fit. 
The data favor a non-zero magnification bias correction 
at the $2\, \sigma$ level. Our estimate for magnification
bias, $A=1$, is 1.5-sigma away from the actual best fit.
Although this is not a very significant deviation, it might
hint that the real lensing effect is larger that the prediction
in the Appendix. This could be caused by a slope $s>1.5$ or
by the combination of non-linear redshift space distortions and
non-linear bias, which are hard to model.

We should also mention that a simple $\xi = 0$ model 
carries a $\Delta \chi^2 = 6$, and so it is only slightly disfavored
compared to the best-fit model. However, a $\xi = 0$ model
is clearly not a good description of the data once we include smaller scales.

The radial shift parameter $D_r$ is constrained to be:
\begin{eqnarray}
D_r = 0.998 \pm 0.037 \quad (1 \sigma)
\label{eq:dr}
\end{eqnarray}
when we marginalize over the other parameters.
The best fit value of $D_r$ is very close to our
fiducial LCDM model with $\Omega_m=0.25$. The $\pm 1$-sigma range (i.e. models with 
$D_r=1.035$ and $D_r=0.961$)
are shown in Fig.\ref{fig:losdirectionmodel2}
as blue dot-dashed lines around the best fit model (black solid line). The
corresponding value of $H(z)$ is shown in the top entry of Table \ref{tab:rbao2}.
In principle one
could find stronger constraints in $D_r$ by fitting to smaller $\pi$ scales, but this
will be sensitive to the modeling of nonlinear redshift space distortions (i.e. finger-of-god).

\begin{figure}
\centering{ \epsfysize=7.5cm\epsfbox{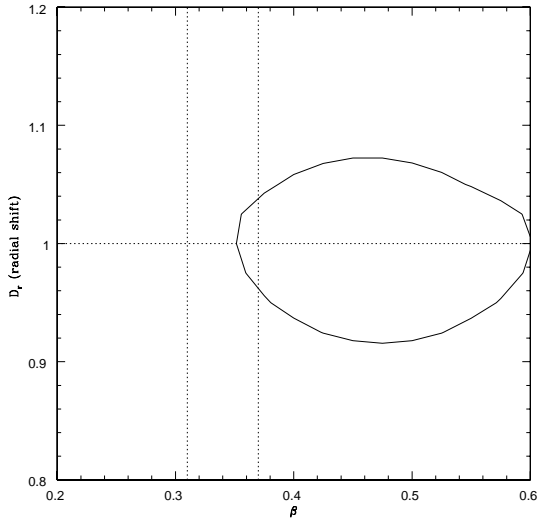}}
\caption{
Joint constraints on the radial shift parameter $D_r$ and $\beta$
when no prior is assumed on $\beta$.
The solid curve shows the $\Delta\chi^2 = 1$ contour.
The vertical lines delimit the $1-\sigma$ region on $\beta$ from 
the observed quadrupole and monopole (Paper I).
\label{fig:chi2betaD}}
\end{figure}

Next, we investigate the robustness of our model fits.
We have systematically explored the constraints on $D_r$
when the priors are relaxed. An example is shown in
Fig. \ref{fig:chi2betaD}. Here, we remove the prior on $\beta$.
The central value for $D_r$ remains remarkably robust. The errorbar
on $D_r$ does increase, as expected, but it is not unduly large.
The best-fit $\beta$ from the LOS data (with no prior on $\beta$) 
lies outside the $1-\sigma$ region from the quadrupole and monopole data, but
the errorbar from LOS data alone is quite large and there is overlap 
between constraints from the two different kinds of data.

We have also investigated what happens if we fit to a different range of scales
$\pi = 70 - 140$ Mpc/h. The best fit model has $\chi^2 = 8$ for 8 degrees
of freedom. A model with no BAO has a $\Delta \chi^2 = 10$ from the best fit,
and a model with no magnification bias has a $\Delta \chi^2 = 4$ for the 
best fit. The results are therefore quite similar to those from $\pi = 40 - 140$ Mpc/h,
but the errors on $D_r$ in Eq.\ref{eq:dr} increase by almost a factor of 2 for  
$\pi = 70 - 140$ Mpc/h, indicating that the $H(z)$ constraint is partially driven by the shape of 
the correlation rather than by just the position of the radial BAO peak. We will show in next
section how to measurement the BAO peak position with independence of the shape, which will provide
an absolute measurement of $H(z)$ (independent of h).

\begin{figure}
\centering{ \epsfysize=7.5cm\epsfbox{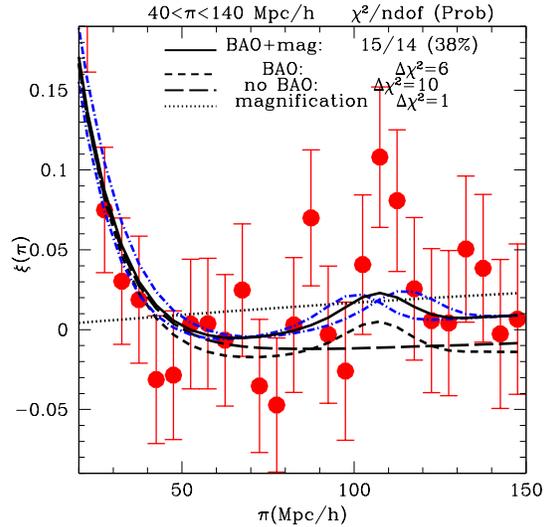}}
\caption{
Same as Fig.\ref{fig:losdirectionmodel2} for $z=0.15-0.30$.
\label{fig:losdirectionmodel3}}
\end{figure}

Fig.\ref{fig:losdirectionmodel3} shows the  same analysis for the nearby redshift
slice $z=0.15-0.30$ which has lower density and much larger errorbars. 
The qualitative results are nevertheless
similar: the model fits the data well, the no-BAO model is ruled out with $\Delta \chi^2=10$
and the model without magnification is ruled out to $\Delta \chi^2=6$. 
The value of $D_r$ is close to unity but constraints on $D_r$ increase to about $7\%$
as indicated by the dot-dashed lines in the figure. 


For a global view of how well the model and the data agree, see
Fig.10 in paper I, where we plot
the data plus model in the $\sigma-\pi$ plane
for the full sample. Visually, it looks quite good. Doing a detailed
overall fit to $\xisp$ is complicated in practice because of the small scale
modeling \citep[eg see][]{paper2}. We leave this task for a future analysis
and focus here in the line of sight data.

\subsubsection{The BAO Feature}
\label{sec:BAOfeature}

\begin{figure*}
 \centering{\epsfysize=4.2cm\epsfbox{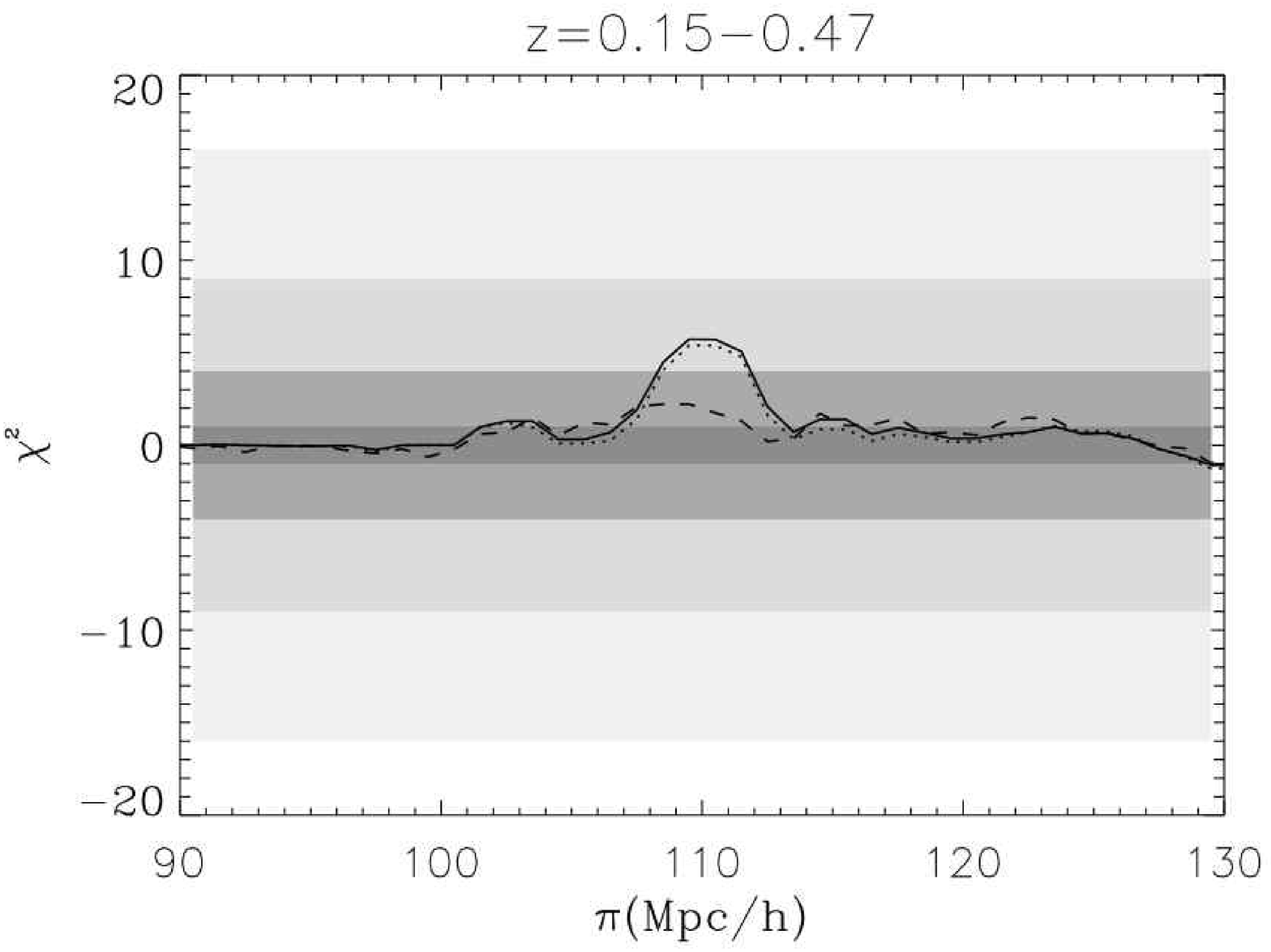}}
 \centering{\epsfysize=4.2cm\epsfbox{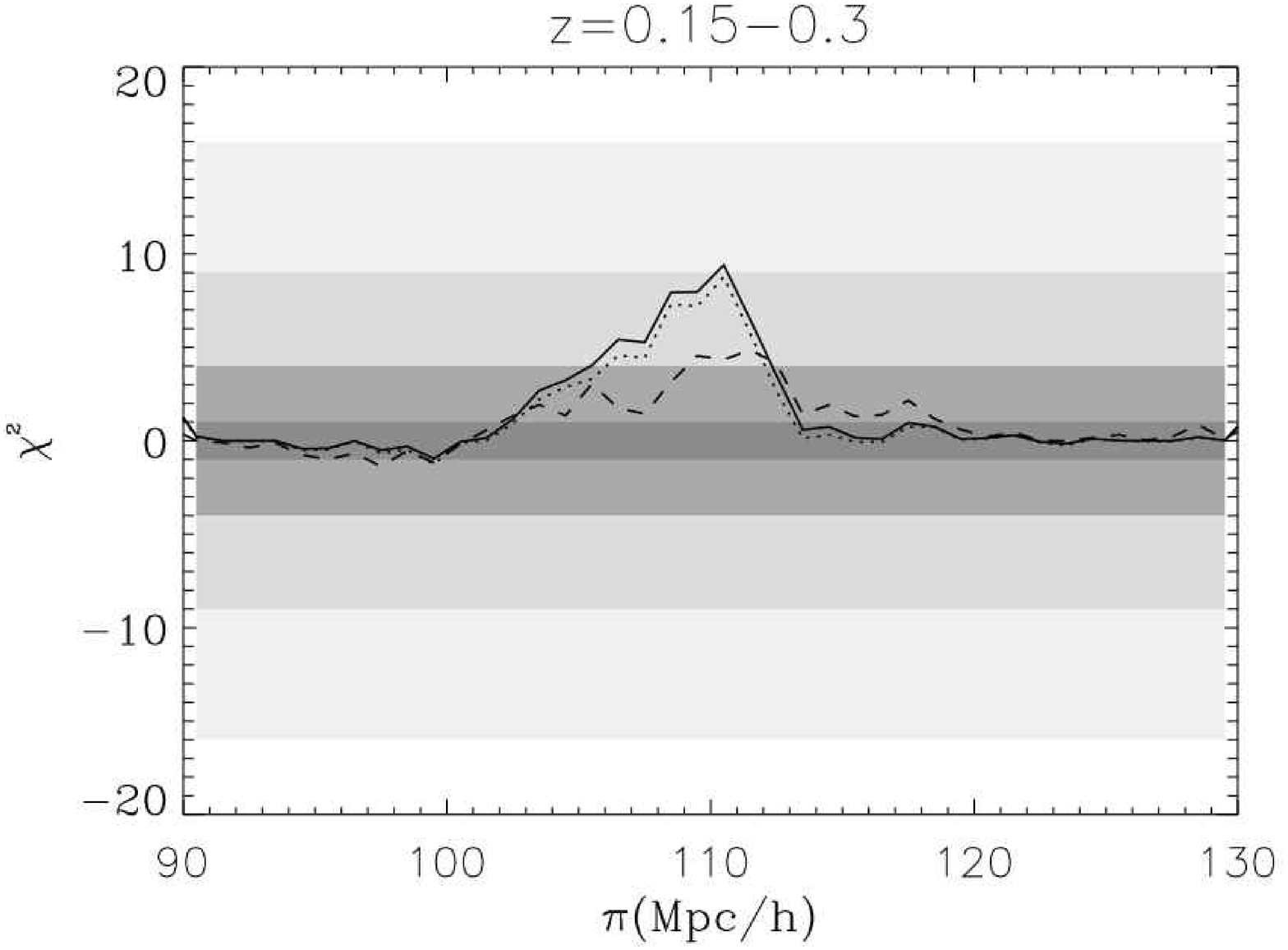}}
\centering{ \epsfysize=4.2cm\epsfbox{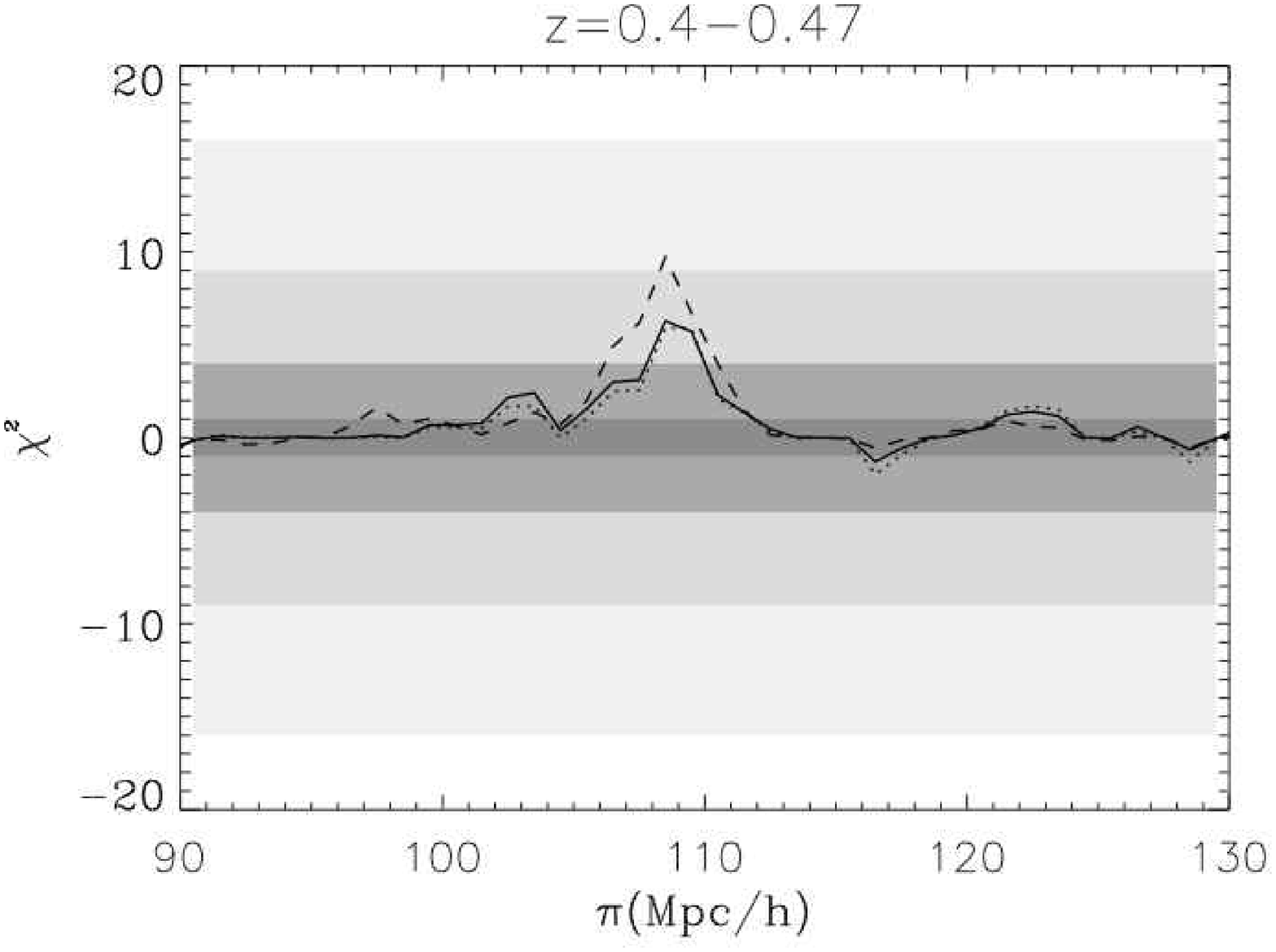}}
  \caption{In this figure we plot the significance of the detection (away from zero)
as $\chi^2=(S/N)^2$
without covariance (solid line) and including the covariance (dotted line)
for bins of 5 Mpc/h (incrementing in steps of 1 Mpc/h).
We also plot the results for a more conservative mask and a smoother selection function (dashed
lines). Each panel corresponds to a different redshift slice, as labeled in the figures. 
The gray zones indicate a 1-$\sigma$, 2-$\sigma$, 3-$\sigma$ or 4-$\sigma$ detection 
 \label{fig:baos2nall}}
\end{figure*}

Next, we would like to focus on the BAO feature in the LOS around $110$ Mpc/h, and 
assess its statistical significance without recourse to model fitting like
in the last section.
We compute the $\chi^2$ against a null signal i.e. $\chi^2 = (S/N)^2$ where
$S$ is simply the observed $\xi$ itself.
The result is plotted in Fig.\ref{fig:baos2nall} where we show results for
the full sample and for the low and high redshift slices. 
Dotted lines
use the full covariance in Fig.\ref{fig:covar}, while solid lines neglect the covariance 
\footnote{
Suppose we label the pixels by $i$. At pixel $i$, the $\chi^2$ neglecting
covariance is $\xi_i^2 / C_{ii}$, and the $\chi^2$ including covariance is
$\sum_j \xi_i C^{-1}_{ij} \xi_j$. Here, $\xi_i$ is the signal at pixel $i$,
and $C_{ij}$ is the error covariance between pixel $i$ and pixel $j$, and
$C_{ii}$ is simply the diagonal variance at pixel $i$.
}.
The difference is small reflecting a small covariance in the radial
distance. The bin width is 5 Mpc/h, so there are 2-3 independent measurements
across the BAO peak.
A peak is clearly detected except from the intermediate 
redshift slice, z=0.3-0.4, which is not plotted here ($\chi^2$ just
fluctuates inside the 2-sigma region).
Errors are bigger in this  z=0.3-0.4 slice because there are fewer number of pairs 
in the radial direction at these scales.
This sub-sample also has a lower galaxy bias because it includes
less luminous galaxies (see Paper I), hence a lower signal and, consequently, less
signal-to-noise (recall we are shot noise dominated).

To assess the effects of possible systematics,
we have repeated the analysis using an angular mask with 
10-20\% less area (and fewer
galaxies), which are safely inside ``good'' plates, 
and with a radial selection function which is smoother (dz=0.05)
than our default value of dz=0.02 (see Appendix in Paper I).
The results are shown as dashed lines in Fig.\ref{fig:baos2nall}.
The BAO peak remains significant in both the low and high redshift slices,
but less so for the full sample.
We should emphasize, however, this last statement on the full sample
only pertains specifically to the $(S/N)^2$ right around
the BAO feature at $110$ Mpc/h.
When we carry out a model fitting to the full sample over a wider range in scales
(e.g. $40 - 140$ Mpc/h), we find that a no-BAO model is in fact
always disfavored by more than $3 \sigma$, regardless of
the mask or radial selection function (see \S \ref{sec:BAOmodel} and Fig.\ref{fig:losdirectionmodel2}).


\section{A Direct Determination of $H(z)$}\label{sec:interpretation}

Our discussions above suggest
we could attempt to infer $H(z)$ from the LOS data in  two ways.
One is to carry out a model fitting as in \S \ref{sec:BAOmodel}.
Constraints on the shift parameter $D_r$ can be translated directly
into constraints on $H(z)$. The result we find is shown as the
first entry in Table \ref{tab:rbao2}. The statistical error comes from
the marginalization explained in the above section, while the systematic
error correspond to the difference in $D_r$ when we use the different versions
of the data shown in Fig.\ref{fig:LOSsys}.
We call this the ``Shape Method''.
Note that this method gives $H(z)$ in units of $H_0$ and it
relies on not just the BAO feature around $110$ Mpc/h,
but also the shape of the correlation function on smaller scales.
Such a method is very analogous to how $(D_A^2 / H)^{1/3}$ is inferred
from the monopole data in the discovery paper by \citet{detection}.

\subsection{The Peak Method}
 \label{sec:BAOfeature2}

The second method to measure $H(z)$ is less model dependent and focus entirely
on the BAO feature around $110$ Mpc/h. We find the peak location,
and use that as a standard ruler to measure the radial distance. 
We call this the ``Peak Method''.
We will compare both methods in this section. Note that in the shape method we
will only constraint $H(z)/H_0$ while in the peak method we will get
$H(z)$ directly.

Operationally, for the peak method, we use the relation:
\begin{equation}\label{eq:obtentionh}
H(z)_{\rm true}={ r_{\rm BAO} \over r_{\rm WMAP}} H(z)_{\rm ref}
\end{equation}
where $r_{\rm WMAP} = 153.3 \pm 2.0$ Mpc (Table 3 in \cite{wmap5}) is the comoving acoustic scale inferred
the cosmic microwave background (WMAP5), and $r_{\rm BAO}$ is the apparent BAO scale inferred from data using the 
fiducial expansion rate
$H(z)_{\rm ref} = H_0\sqrt{0.25(1+z)^3 + 0.75}$ to convert redshifts to distances.

\subsection{Statistical Error}
\label{sec:staterror}

\begin{table}
\begin{center}
 \begin{tabular}{ccc}
\hline
 Sample  &   $r_{\rm BAO} \pm \sigma_{st} \pm \sigma_{sys}$ &     \\
 z range  (mean)      & Mpc/h    & $\Delta z_{\rm BAO}  \pm  \sigma_{st}  \pm  \sigma_{sys}$ \\
\hline
\hline
 0.15-0.30 (0.24) & $110.3 \pm 2.9 \pm 1.8$ & $0.0407 \pm 0.0011 \pm 0.0007$\\
 0.15-0.47 (0.34) & $110.5 \pm 3.6 \pm 2.1$ & $0.0428 \pm 0.0014 \pm 0.0008$\\
 0.40-0.47 (0.43) & $108.9 \pm 3.9 \pm 2.1$ & $0.0442 \pm 0.0015 \pm 0.0009$\\
\hline
\end{tabular}
\caption{The BAO fiducial scale $r_{\rm BAO}$ in the LOS direction calculated
with a flat reference $H_{\rm ref}(z)$ cosmology
of $\Omega_m=0.25$, for three redshift slices: in parenthesis 
is the respective pair-weighted mean redshift,
and $\sigma_{st}$ and $\sigma_{sys}$ are the statistical and systematic errors on $r_{\rm BAO}$.
The direct $\Delta z_{\rm BAO}$ measurement is shown in the third column, which relates to
the fiducial scale as $\Delta z_{\rm BAO} = r_{\rm BAO} H_{\rm ref}(z)/c$
and is independent of the value chosen for $H_{\rm ref}(z)$.
\label{tab:rbao}
}
\end{center}
\end{table}


\begin{figure}
\centering{\epsfysize=7.cm\epsfbox{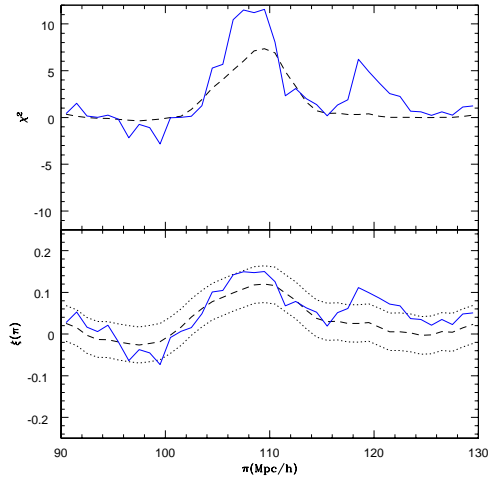}}
\caption{{\it Lower panel:} dashed line and dotted lines show the mean
and dispersion for a model of the correlation function in the radial direction
for the redshift slice z=0.15-0.30. This model uses a smoothed version of the
real data as the baseline (the mean).
The continuous line shows one of the $10^4$ Monte Carlo realizations, which have
noise added in a way that accounts properly for covariance seen in the data.
{\it Top panel:} $\chi^2=(S/N)^2$ in the model (dashed line)
and in the same one Monte Carlo realization (continuous line). 
\label{fig:peak}}
\end{figure}


The peak measured in the radial direction is broad, just as expected
from the BAO signal (see Fig.\ref{fig:losdirectionmodel2}). 
We will take the BAO position
to correspond to the location of the maximum in the (S/N) around the broad BAO peak,
and we will use Monte Carlo simulations to estimate the associated errorbar.
The resulting fiducial BAO scale location  $r_{\rm BAO}$ are shown  in Table \ref{tab:rbao}. 

The Monte Carlo simulations are generated as follows. 
We model the signal as a smoothed version of the data. We use
a top-hat smoothing of 5 Mpc/h width and radial bins centered in 1Mpc/h 
intervals. Such a model
is shown as a dashed line in Fig.\ref{fig:peak} for the sample z=0.15-0.30.
Given the model, we then add noise with the same covariance and 
binning as in the real data. A detailed discussion of our noise model, and
the tests we have applied to it, is given
in \S \ref{estimateError}.
Note that while we have chosen pixels of $5$ Mpc/h in size,
the data actually contain information on finer resolution. In our analysis
and in our simulations, the pixels are in fact incremented in the $\pi$ direction
by steps finer than $5$ Mpc/h to facilitate the search for the BAO peak, and
the full covariance between these overlapping (and non-overlapping) pixels
is taken into account.

\begin{figure}
\centering{\epsfysize=7.cm\epsfbox{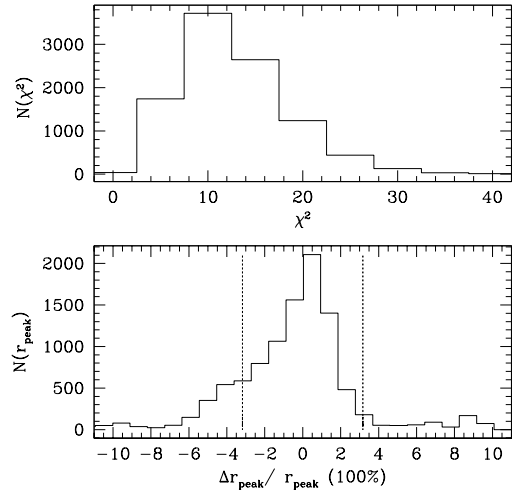}}
\caption{{\it Lower panel:} Distribution of relative values
in the BAO peak location in $10^4$ Monte Carlo simulations, taking a smoothed
version of the data as the mean
for the redshift slice z=0.15-0.30.
 {\it Top panel:} Distribution of peak values of $\chi^2$. Here,
$\chi^2 = (S/N)^2$ with $S$ being the difference between $\xi$ and zero i.e. 
this is $\chi^2$ against null detection.
\label{fig:Ppeak}}
\end{figure}

Fig.\ref{fig:peak} shows one realization
of such simulations. For each realization, we
infer the BAO peak location using exactly the same method as applied to real data.
We repeat this $10^4$ times and estimate the
distribution of the inferred BAO peak location. 
The errorbar is 
given by the difference between the inferred peak location
and the input BAO scale in the model. 
The distribution of fractional error in the BAO location, and the distribution of
peak $\chi^2=(S/N)^2$ values, are shown in  Fig.\ref{fig:Ppeak} for the
same $z=0.15-0.30$ sample. We find an rms
fractional error between 2.6 and 3.5$\%$, depending on the redshift sample used. 
The resulting errorbars in the peak location 
are shown as $\sigma_{st}$, the statistical error, in Table \ref{tab:rbao}.

\subsection{Systematic Error}

There are two sources of systematic errors. One arises from the measurement
process, and the other is theoretical in nature. Let's discuss the theoretical
one first. Recently, \citet{baughbao} undertook a thorough investigation of
systematic effects in the determination of the sound horizon from the galaxy
correlation function. Their Fig. 2 shows that at $\Omega_m = 0.25$, the peak
in the correlation function systematically underestimates the 
true sound horizon by about 1.5\% (this error
increases to 2\% for $\Omega_m=0.2$ and reduces to 1\% for $\Omega_m=0.3$).
One can understand this result by modeling the monopole correlation as a sum of
a power-law and an Gaussian peak center on the BAO position. As we change the
power-law index and the relative amplitude of the peak the maximum of the
correlation can shift with respect to the maximum of the Gaussian. This effect should be
different (and weaker) for the LOS, because the underlaying correlation
is flatter (and in fact slightly increasing with scale).
On the other hand, magnification bias tends to move the peak by a similar
amount in the opposite direction \citep{hui1}. 
Correcting for these biases is in principle possible, but requires accurate LOS modeling
which is not currently feasible.

A larger source of systematic
uncertainty comes from the model used in the Monte Carlo simulations
in the previous section. If instead of the data or a smoothed version of
the data, we take our best fit theoretical model in section 
\ref{sec:BAOmodel} as the mean input to the Monte Carlo noise realizations 
we find a much larger error in the BAO position, of size $4-8\%$ depending
on the different assumptions.
The reason for this is that the model has a lower amplitude than the data
around the peak
and models are quite degenerate to uncertainties in the cosmological
parameters. Even when this is not a statistically 
significant departure, the smaller amplitude in the model results in 
a large degradation in  the recovered  error in the peak position.
We can of course choose models that have higher peak amplitude, and that
are still allowed by the data. Those models will provide artificially small
errorbars.  

\begin{figure}
\centering{\epsfysize=7,cm\epsfbox{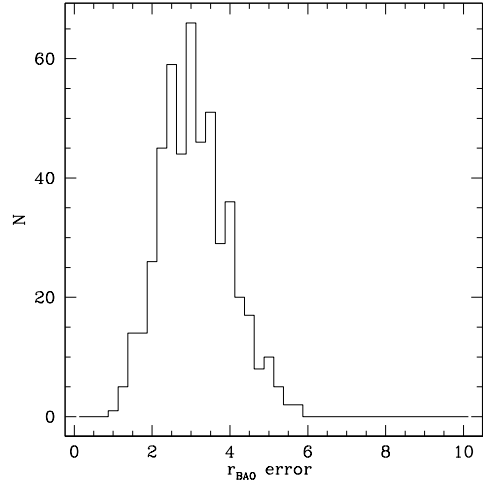}}
\caption{Distribution of errors in the peak positions using 500 different
baseline models (each model being a realization of the data plus noise). For
each baseline model we do $10^4$ montecarlo simulations to estimate the
corresponding distribution and error (ie as in lower panel of Fig.\ref{fig:peak}).
The variance in this distribution gives us an estimate of the error in the error.
This is for the redshift slice z=0.15-0.30.
\label{fig:errorerror}}
\end{figure}

We  can account for this systematic deviations in a model independent way
by noticing that the observations must in fact result from the true model plus 
a realization of the noise. Thus, if we add noise to the observations we
will in fact explore, by definition, all possible models that can agree with the data.
We can use these new sets of models (data+noise realizations) as a new baseline to our 
Monte Carlo simulations and explore how much different the error can be
when we allow for a variation in the baseline. We have run 500 such data+noise realizations
and have run $10^4$ Monte Carlo simulations for each of them (as in previous
section) to find the distribution of possible errors. 
The result is shown in Fig. \ref{fig:errorerror} for the z=0.15-0.30 slice.
The variance in the resulting distribution provides us with an estimate of the
systematic uncertainty in our error determination.

\begin{table}
\begin{center}
 \begin{tabular}{|cc|c|}
\hline
Shape Method & & \\
\hline
redshift  & $z_m$  ~~& $\,\, H(z)   \, \pm \, \sigma_{st}  \, \pm \, \sigma_{sys} \,\,$  \\
range &  &  km/s/Mpc $h_{72}$   \\
\hline
 0.15-0.47 & 0.34 ~~&  $\,\, 83.87  \, \pm \, 3.10 \, \pm \, 0.84 \,\,$\\
\hline
\hline
Peak Method & &\\
\hline
redshift & $z_m$  ~~& $\,\, H(z)   \, \pm \, \sigma_{st}  \, \pm \, \sigma_{sys} \,$  \\
range &  &   km/s/Mpc   \\
\hline
 0.15-0.30 & 0.24 ~~& $\,\, 79.69 \, \pm \, 2.32 \, \pm  \, 1.29 \,\,$\\
 0.15-0.47 & 0.34 ~~&  $\,\, 83.80 \, \pm  \, 2.96 \, \pm  \, 1.59 \,\,$ \\
 0.40-0.47 & 0.43 ~~&  $\,\ 86.45 \, \pm \, 3.27 \, \pm \, 1.69  \,\,$\\
\hline
\end{tabular}
\caption{
The inferred $H(z)$ with its associated errors, ie statistical and systematical, 
 for each redshift slice. The top entry corresponds
to fitting the shape of the LOS correlation to models and marginalized 
over cosmological parameters
(note that this is in units of $H_0=72 h_{72}$ Km/s/Mpc). Bottom values use the
Peak Method explained in Section \ref{sec:BAOfeature2}, which is model independent. \label{tab:rbao2}
}
\end{center}
\end{table}

The statistical $\sigma_{st}$ error is taken
to be the variance when we use just the data as a based model rather than the mean
in the distribution of errors from data+noise, which is larger because the error
has been added twice.

An additional source of our systematic error is estimated by taking
the difference between results obtained using different angular masks
and selection functions (see Fig. \ref{fig:LOSsys} and associated
text).  Note that our results do not change significantly when we use
different masks in DR6 or even the DR7 data, which became available
after the first version of this paper.  This is then added to the
above systematic error to produce $\sigma_{sys}$ given in Table
\ref{tab:rbao}.  Errors are then propagated to $H(z)$ in Table
\ref{tab:rbao2}.

One additional source of systematic error might have occurred to the
reader.  Our estimate of the BAO scale is not strictly from the LOS
direction: as discussed in \S \ref{radialpeak}, we have used an array
of pixels extending in the $\pi$ direction but with $\sigma = 0.5$
Mpc/h to $5.5$ Mpc/h, instead of being centered at $\sigma = 0$ (to
avoid the fiber collision zone).  This amounts to approximating the
BAO scale $r = \sqrt{\pi^2 + \sigma^2} \sim \pi$.  The lowest order
fractional correction is $(\sigma/\pi)^2/2 \sim 4 \times 10^{-4}$
which is negligible.

\subsection{Results on $H(z)$}

\begin{figure}
\centering{ \epsfysize=7.cm\epsfbox{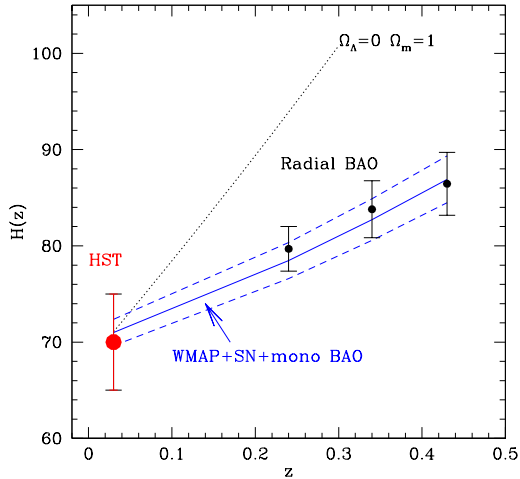}} 
  \caption{$H(z)$ obtained using the ``Peak Method'' over  three different redshift slices 
 (small dots with errorbars) compared to the HST value at $z\simeq 0.03$
 (big dot with errorbar). We also plot for comparison
the best value for H(z) from WMAP5 + SNIa + BAO monopole (solid line) and 
its associated allowed region at 2-sigma level
(dashed lines), assuming a flat w=-1 LCDM cosmology. For reference, the
two dotted lines show the values of $H(z)$ in open and closed
models without cosmological constant.
 \label{fig:hchange}}
\end{figure}

 Table \ref{tab:rbao2} summarizes the different $H(z)$ estimations
and corresponding errors. The values of 
$H(z)$ from the ``Peak Method'', ie 
using Eq.(\ref{eq:obtentionh}), for the low (z=0.24) and high (z=0.43) redshift slices
are independent of each other because they come from galaxies that
are far apart. These values are the main result of this paper, as they are both model
and scale independent (see previous section). 
In contrast, the values from the ``Shape method'' have been
shown to depend on the range of scales used in the fit. 
 The main virtue of this ``shape method'' is to show that
the LOS correlation agrees well with the BAO prediction, which
indicates that measured peak does in fact correspond to the BAO.

The values of $H(z)$ from the peak method are independent of and in excellent agreement with
the predictions for H(z) according to the current combined constraints from
WMAP5, supernovae Ia and the monopole BAO \cite{wmap5}, assuming a flat
LCDM  cosmology (ie w=-1). In Fig.\ref{fig:hchange} we compare our
estimates with $H_0=70 \pm 5$  km/s/Mpc  from  HST \cite{HST2007}  at $z \simeq 0.03$
and with the $H(z)$ inferred from WMAP5 modeling with  $H_0=71.9 \pm 2.6$ km/s/Mpc 
and $\Omega_m=0.258 \pm0.030$ for a flat  universe LCDM (w=-1) \cite{wmap5}.
For a flat cosmological constant dominated model, our measurements extrapolate to 
\begin{equation}\label{eq:H0}
H_0=71.83 \pm 1.55 ~(\pm 1.03) \;{\,\rm km/s/Mpc}
\end{equation}
if we use $\Omega_m=0.245 \pm 0.020$ 
obtained from fitting $\xisp$ to the same data in  Paper I.\\

\begin{figure}
\centering{ \epsfysize=6.5cm\epsfbox{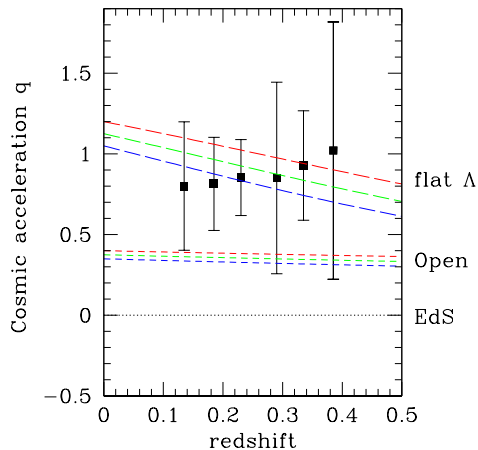}} 
  \caption{Estimates for cosmic acceleration $q(z)$
based on the different pairs of $H(z)$ measurements shown
in Fig.\ref{fig:hchange}.
Dashed color lines show values for flat  (top lines) and open models 
with $\Omega_m$=0.20 (red), 0.25 (green) and
0.30 (blue). Dotted line correspond to EdS model ($\Omega_m=1.0$ and $\Omega_{\Lambda}=0$).
 \label{fig:qchange}}
\end{figure}

We can estimate the acceleration parameter 
\begin{equation}\label{eq:defqz}
q[z]  \equiv {d\ln H/{d\ln a}} +3/2,
\end{equation}
defined to give $q=0$ for the EdS (flat universe without Lambda) Cosmology.
Using our two independent values of $H(z)$ at $z=0.24$ and $z=0.43$, 
we find  at intermediate redshift, z=0.34:

\begin{equation}\label{eq:qz}
q[z \simeq 0.34] = 0.93 \pm 0.33
\end{equation}

If instead we compare our value for the full sample $z=0.15-0.47$
at mean $z=0.34$ with the HST value $H_0=70 \pm 5$  km/s/Mpc at $z \simeq 0.03$ we find:

\begin{equation}\label{eq:qz2}
q[z \simeq 0.18]= 0.82 \pm 0.30
\end{equation}

Both results provide independent evidence for cosmic acceleration
(ie away from EdS model).
The mean value seems to increase with redshift which is contrary to
all expectations (cosmic acceleration should turn into deceleration at
high redshifts) , but this trend is not significant given the errorbars.
The interest of these results is that for the first time we have direct
(geometrical) measurements of cosmic acceleration at different redshifts.
Fig.\ref{fig:qchange} shows the different estimations of $q(z)$ for the
6 possible combinations of the 4 values of $H(z)$. Note that these
estimates are not independent. Current errors are larger than typical evolution
expected in this redshift ranged for different cosmological
models (shown as dashed lines) so it is hard to
conclude more with this data.

\section{Discussion \& Conclusions}\label{sec:conclusions}

In this paper, we have studied in detail a peak in the LRG correlation function,
in the monopole, in the circular ring on the $\sigma-\pi$ plane, 
and especially in the LOS direction. 
Its location around $110$ Mpc/h is consistent with the interpretation that
it originates from baryon acoustic oscillations.
Its significance can be assessed in several ways. 
We have performed a parametric fit to the monopole: the data requires
a non-zero $\Omega_b$ with high significance. In fact, we find that data
appears to favor a value that is slightly higher than the standard WMAP5 value.
However, the difference is less than $2-\sigma$ at both the full sample
and the low redshift slice
$z = 0.15 - 0.30$, but is larger at the high redshift slice $z = 0.40 - 0.47$
(see Fig. \ref{fig:fit.slice}). 
We have also found this tendency to high $\Omega_b$ 
in paper III \citep{paper3} of this series, where we analyze the BAO 
scale in the  3-point correlation function, which tests non-linear growth of 
perturbations, in contrast to the 2-point correlation function, that tests linear perturbations.
Plots of the signal and the signal-to-noise of $\xi$ in the $\sigma-\pi$ plane
(Fig. \ref{fig:pisigma} and \ref{fig:s2nps.11}) provide a reassuring view
of the reality of the signal: a negative valley (in blue) of negative correlations
at $\pi \sim 50 - 90$ Mpc/h together with a ring (in red) of positive correlations
at $\sim 110$ Mpc/h. 
The BAO peak in the LOS direction is pronounced, and detected
with significance (Fig. \ref{fig:losdirectionmodel2} and \ref{fig:baos2nall}). 

It would be hard to explain such a peak in the 2-point monopole, 
in the plane $\sigma-\pi$, in the LOS and in the 3-point function with 
systematic effects. We have checked for possible systematics by varying the angular mask and
the radial selection function, and find the detection robust using either DR6 or DR7 SDSS data
(see \S \ref{sec:SystematicsTests}).
The data on scales $\pi = 40 - 140$ Mpc/h is well fit by a model that includes 
linear redshift space distortions, magnification bias and BAO.
In fact, a model with no BAO is disfavored by the data at the $3 \sigma$ level,
while a model with no magnification bias is disfavored at the $2 \sigma$ level.
Our detection of the BAO modulation in the LOS direction 
is helped by the combination of two effects:
redshift space distortions make $\xi$ negative on scales of $50 - 90$ Mpc/h
while magnification bias gives a significant positive boost on larger scales.
%

We thus have significant evidence that the LOS correlation 
on scales $40 <\pi <140$ Mpc/h reproduces well the expected BAO signal. 
Consequently we use the LOS data to infer $H(z)$ in two ways which we have labeled
the shape and the peak method. In the shape method 
we have used the shape of the LOS correlation to fit $H(z)/H_0$ marginalized over other
cosmological parameters. The result is shown as the first entry to Table \ref{tab:rbao2}.
In the peak method we find the location of the peak position and use it as a standard ruler
using Eq.\ref{eq:obtentionh}.
This is a more direct, geometric test in the sense that the BAO is used
strictly as a standard ruler (i.e. we are not using model-dependent shape of the correlation function), 
and in the sense that we constrain $H(z)$ rather than integrals thereof.
The results are shown as 3 bottom entries of Table \ref{tab:rbao2}.


There are two parts to our error analysis. One is the errors on $\xi(\sigma, \pi)$ itself,
and the other is the errors on our measurement of the peak location.
Both have been extensively tested with simulations.
The error model for $\xi$ is obtained from what is to our knowledge
the largest ever cosmological simulation 
run to date, MICEL7680 with 453 ${\,\rm Gpc^3/h^3}$, $2048^3$ dark matter particles
and 107 million halos in a single box \citep{mice,mice2}. We are able to 
create out of this simulation 216 independent mock LRG catalogs, each of which 
has a similar size to our SDSS DR6 sample. Our mock catalogs have 
the same number density and very similar 2- and 3-point
functions compared to data. Fig.\ref{fig:xigmu3} shows, for the first time on BAO scales, 
a comparison of the redshift space LOS linear model with non-linear measurements from 
simulations. 
Both model and simulation show very similar shapes and a prominent BAO peak 
which validates both our modeling and our simulated mocks.
The error model constructed from these simulations is further validated by
a comparison with jack-knife errors obtained from the data itself.
Further details can be found in \S \ref{sec:nonstandard} and Fig.\ref{fig:errorvalid2}.

The other part of our error analysis consists of using Monte Carlo simulations
to simulate our peak measurement process. This is described in \S \ref{sec:staterror}.
This method involves a further systematic uncertainty: because the data is quite noisy
we can not be certain of what is the true model for the mean of the Monte Carlo
simulations. We have used the data itself (or a smooth version of it) to
measure both the statistical and systematic errors on these measurements. 
Thus, these errors are more
model independent than the ones in the shape method, 
which depend strongly on the range of scales used for the fit.

\begin{figure}
\centering{ \epsfysize=6.5cm\epsfbox{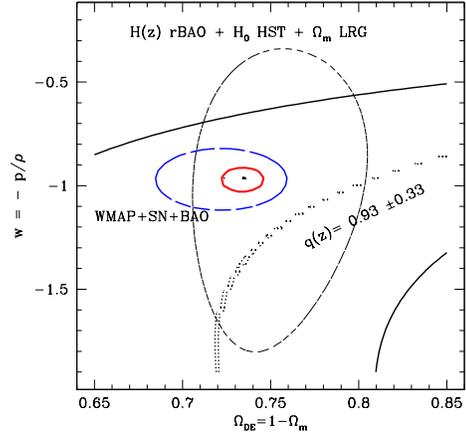}}
  \caption{
The dotted line and continuous (black) open contours
across the figure show the best fit and 1-sigma contour from $q[z=0.34]$ in Eq.\ref{eq:qz}, 
ie using only
our two independent $H(z)$ measurements. Long-dashed 
line contour show $\Delta\chi^2=6.18$ (2-sigma for 2 dof)
 constraints from combining our $H(z)$ measurements with
$H_0$ from HST and our $\Omega_m$ estimate from redshifts distortions (Paper I).
Short-dashed contours (blue) show the  corresponding $\Delta\chi^2=6.18$  constraints from 
the external data-set WMAP5, SNIa and BAO monopole. 
The inner red continuous contour shows  $\Delta\chi^2=1$ 
when combining both measurements.
 \label{fig:fitwOl}}
\end{figure}

Given our measurements of $H(z)$, it is natural to ask: what constraint do they
put on the dark energy abundance and equation of state?
Our measurements of $H(z)$ at $z = 0.24$ and $z = 0.43$ by themselves only
weakly constrain $\Omega_{\rm DE}$ and $w$.
The constraint comes essentially from the acceleration parameter $q(z)$ in
Eq.\ref{eq:qz}. The mean and 1-sigma constrains on $\Omega_{\rm DE} - w$ from 
$q(z)$ are shown as  dotted and continuous black open contours 
across Fig.\ref{fig:fitwOl}. These constraints are very weak as compared to the 
constraints from  WMAP5 + SN + monopole BAO \cite{wmap5}, which are shown
for $\Delta\chi^2=6.18$ as a long-dashed (blue) line contour in Fig.\ref{fig:fitwOl}.
The large short-dashed (black) line contour corresponds to the 2-sigma constraints when 
we combine our $H(z)$ measurements with $H_0 = 70 \pm 5$ km/s/Mpc,
based on the HST analysis \cite{HST2007}, and our
best fit $\Omega_m=0.245 \pm 0.020$ using the same LRG data in Paper I.
The 2-sigma contours from the two independent data sets agree well,
which is not trivial specially given the different assumptions and
measurements involved in each data set. Combining both data sets, we find 
\begin{eqnarray}
w &=& -0.957 \pm 0.053 \\
\Omega_{DE} &=& 0.734 \pm 0.023
\end{eqnarray}
which is shown as red continuous contour in Fig.\ref{fig:fitwOl}.
This is in excellent agreement with the cosmological constant model.
The improvement in the error with respect to the 1-sigma WMAP5+SN+BAO value
of $w = -0.97 \pm 0.06$ and is quite modest.
A separate paper  presents the implication on cosmological parameters
when using our radial BAO scale
measurements in a way that is independent of $H_0$
\citep{paper5}.


\begin{figure}
\centering{\epsfysize=6.5cm\epsfbox{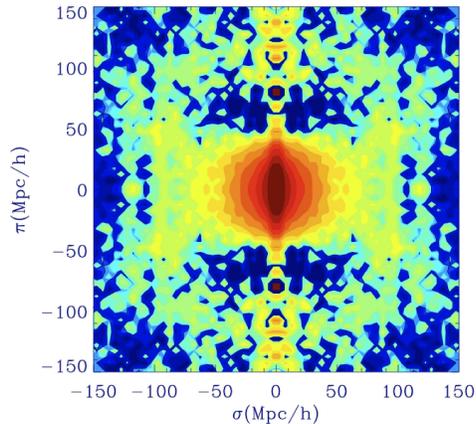}}
  \caption{
$\xips$ in data for the z=0.15-0.30 slice recalculated from
real LRG positions with an added photo-z error of $0.003(1+z)$ as predicted for the
PAU survey. It should be compared with middle panel of Fig.\ref{fig:pisigma}
which uses spectroscopic redshifts.
\label{fig:pisigmaPAUb} }
\end{figure}


To illustrate what might be achievable in the near future,
we show in Fig. \ref{fig:pisigmaPAUb} the expected $\xi(\sigma,\pi)$ 
for the PAU survey \citep{PAU} which has a photo-z accuracy of $0.003(1+z)$.
We redo the analysis using current SDSS data (slice z=0.15-0.30)
but we disperse the LRG positions by this photo-z error 
(we assume a Gaussian distribution with $0.003(1+z)$ dispersion).
The BAO signal does not appear to be washed away, as shown
 in Fig. \ref{fig:pisigmaPAUb}. This is expected 
because the BAO peak is broader than the scale corresponding to the 
photo-z dilution. The PAU Survey
proposes to map over 10 times the SDSS DR6 volume, ie to z=0.9
(this is a factor of 3 in sampling variance error)
with 20 times the LRG number density (ie for $L>L_*$) 
so shot noise will be negligible. All this 
should increase the signal-to-noise over SDSS DR6 by
better than a factor of four.  Thus, there is potential for
a substantial improvement in the near future for the $H(z)$ measurements 
using the techniques presented here.

\section*{Acknowledgments}

We would like to thank Eiichiro Komatsu for pointing out an error
in the value of  $r_{\rm WMAP}$ that we used in the first version of this paper, 
Francisco Castander, Martin Crocce, Pablo Fosalba, Marc Manera and Ramon Miquel
for their advise and support at different stages of this project.
We acknowledge the use of simulations from the MICE consortium 
(www.ice.cat/mice) developed at the MareNostrum supercomputer
(www.bsc.es) and with support from PIC (www.pic.es),
 the Spanish Ministerio de Ciencia
y Tecnologia (MEC), project AYA2006-06341 with
EC-FEDER funding, Consolider-Ingenio CSD2007-00060
and research project 2005SGR00728
from Generalitat de Catalunya. AC acknowledges support
from the DURSI department of the Generalitat de
Catalunya and the European Social Fund.
LH acknowledges support by the DOE grant DE-FG02-92-ER40699
and the Initiatives in Science and Engineering Program
at Columbia University, and thanks 
Shiu-Yeun Cheng and Tai Kai Ng 
at the Hong Kong University of Science and Technology for
hospitality.

\bibliographystyle{mn}
\bibliography{tesi}

\appendix

\section{Numerical Investigations of Nonlinear Effects}
\label{app:nonlinear}

\begin{figure}
\centering{\epsfysize=6.5cm\epsfbox{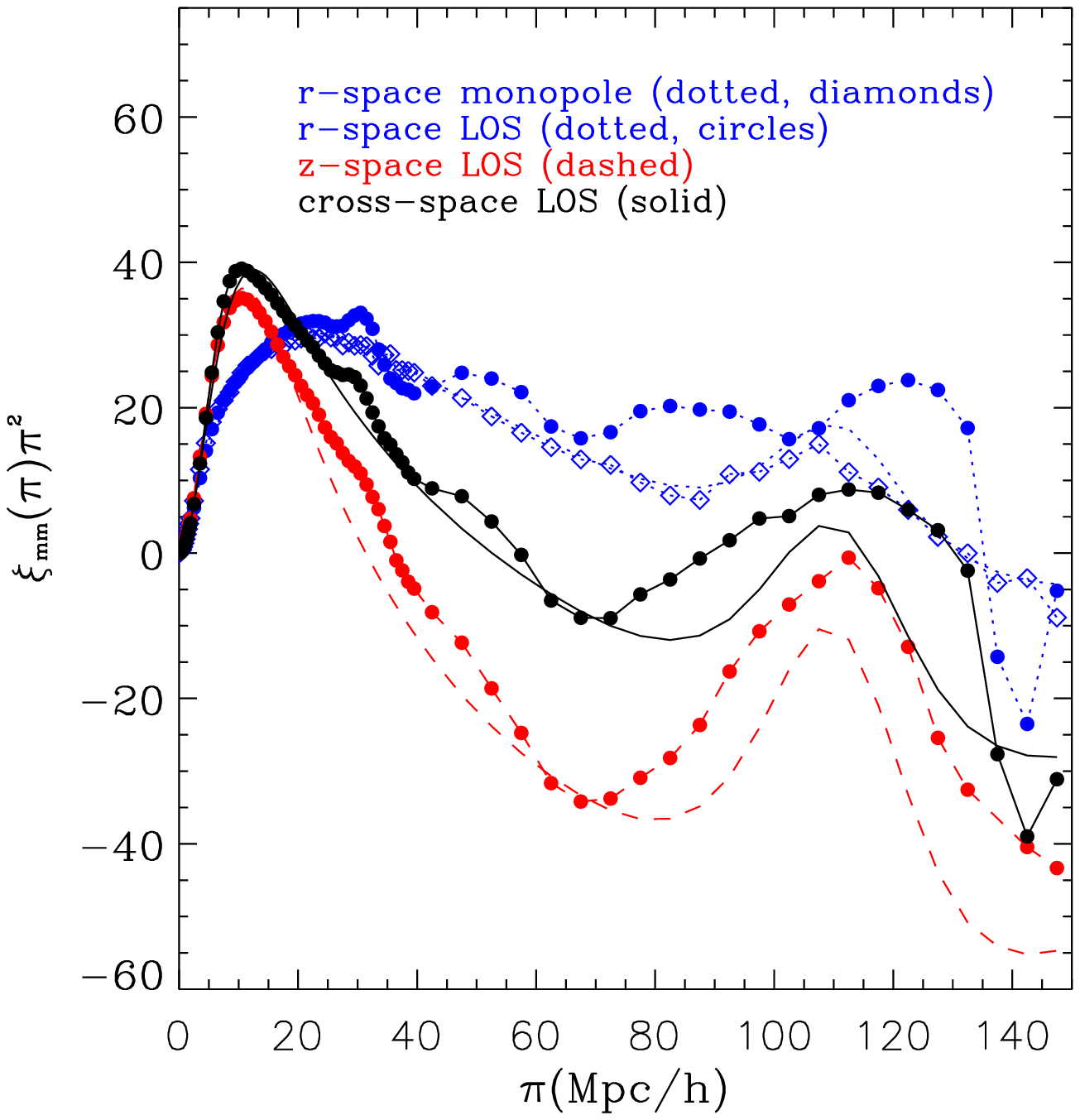}}
\caption{
Line-of-sight (LOS) correlation averaged for
$0<\sigma<5$ Mpc/h. Filled circles connected by lines represent
the following quantities measured from dark matter simulations:
LOS mass autocorrelation in real space (blue dotted), redshift space (red dashed), 
cross real-redshift space (black solid). 
The dotted, dashed and solid lines (with no dots or circles) 
correspond to our analytic predictions for the LOS mass autocorrelation
in real, redshift and cross space.
Finally, also shown is the monopole mass autocorrelation in real space
(unfilled diamonds) which agrees very well
with the prediction and, modulus sampling
variance, should also agree with the LOS mass autocorrelation.
\label{fig:xigmu3}}
\end{figure}

\begin{figure}
\centering{\epsfysize=6.5cm\epsfbox{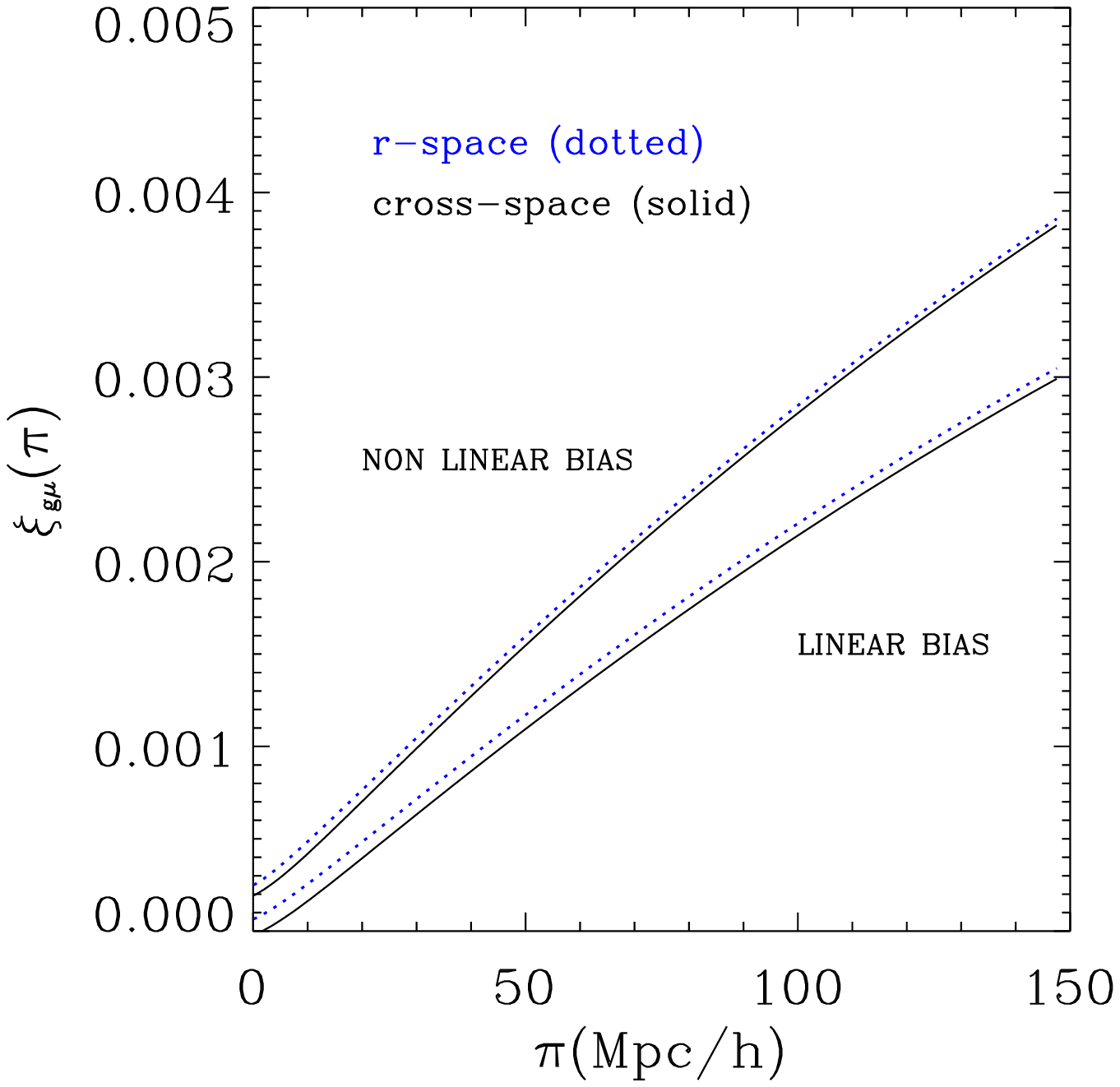}}
\caption{Magnification bias correction based on Eq. (\ref{eq:xigmu3})
(which replaces Eq. [\ref{eq:xigmu}])
for a mean redshift $z=0.34$,  $\Omega_m=0.25$ and  $s=1.5$ for 
linear (bottom lines) or non-linear (upper lines) bias using:
a) the prediction of the cross-correlation $\xi_{gm}$ in real space (dotted),
b) the prediction of the cross-correlation $\xi_{gm}$ in cross space (solid).
Here, cross space means the galaxy is in redshift space and the mass
is in real space. This $\xi_{g\mu}$ is computed with the transverse
separation $\sigma$ averaged over $0 - 5$ Mpc/h.
\label{fig:xigmu} }
\end{figure}

In this Appendix, we describe a number of measurements we have made
to gauge the size of various nonlinear effects on the magnification bias
correction to the galaxy correlation. The measurements are done
on numerical N-body simulations described in the main text.
Recall that the magnification bias correction to galaxy overdensity at
radial distance $\chi_2$ is given by line of sight integral:
\begin{equation}
\delta_\mu(\chi_2) = K \int_0^{\chi_2} d\chi
{(\chi_2-\chi)\chi\over{a(\chi)\chi_2}}~\delta(\chi)
\label{eq:xigmu2}
\end{equation}
where $K=(5 s-2) {3\over 2} H_0^2 \Omega_m$.
Consider now a regrouping of the terms in  Eq.(\ref{eq:deltaobs}):
$\delta_{obs} = \delta_g + \delta_\mu$,
where $\delta_g$ corresponds now to the galaxy overdensity in redshift space
(i.e. including non-linear redshift space distortions). The cross-term contribution
to the observed correlation function  is then 
$\xi_{g\mu} \equiv \langle \delta_g(1)\delta_\mu(2) \rangle$,
where $1$ and $2$ denotes the positions of the
the foreground and background galaxies. 
The LOS separation in redshift space is $\pi$, and
the transverse separation is $\sigma$.
The relevant galaxy-magnification correlation is (replacing
Eq. [\ref{eq:xigmu}]):
\begin{equation}
\xi_{g\mu}^{\chi_2}(\pi,\sigma)= K \int_0^{\chi_2} d\chi
{(\chi_2-\chi)\chi\over{a(\chi)\chi_2}}~\xi_{gm}(\chi_1-\chi,\sigma)
\label{eq:xigmu3}
\end{equation}
where $\xi_{gm}\equiv \langle \delta_g(1)\delta_m(\chi) \rangle$
is the cross correlation of galaxy fluctuations in redshift space with
dark matter fluctuations
in real space. Here, $\chi_1$ is the redshift space radial distance
to galaxy $1$, but $\chi_2$ and $\chi$ are real space distances.
Note  that  $\xi_{gm}$ depends only on the pair separation i.e. 
$(\chi_1 - \chi, \sigma)$,
while $\xi_{g\mu}$ depends both on $\chi_2$ and on the
pair separation $(\pi,\sigma)$.

In our new calculation here we will fix $\chi_2$ to be at
the mean redshift of the survey, 
but we have also studied how the result changes when we take into account 
the fact that 
$\chi_2$ varies across the survey (see below) - basically, if $\chi_1$ are $\chi_2$ are
well separated, $\xi_{g\mu}$ is almost independent of $\chi_2$ (except indirectly
through the LOS separation $\pi$).
One can see how Eq. (\ref{eq:xigmu}) comes about from Eq. (\ref{eq:xigmu3}) --
take the limit in which $\chi_2 - \chi_1 \ll \chi_2, \chi_1$, and ignore
the subtle difference between redshift space and real space distances.
Our task here is to evaluate Eq. (\ref{eq:xigmu3}) exactly using simulations.

First, as a warm-up, we use the dark matter (as opposed to halo) simulations
which have the correct velocities on small scales. 
We compare the mass-mass correlation $\xi_{mm}(\pi,\sigma)$ in real space, redshift space and the cross-correlation 
between real and redshift space. (Note: $\xi_{mm}$ can be thought of as a first proxy to $\xi_{gm}$; we
will examine $\xi_{\rm halo, m}$ below as a better proxy.)
We smooth $\xi_{mm}(\pi,\sigma)$ in bins of 5Mpc/h, as we do with real LRG galaxies. 
In  Fig.\ref{fig:xigmu3}  we show a set of N-body measurements (shown as 
dots connected by lines) for the dark matter
LOS auto-correlation in real space (dotted blue line), 
redshift space (dashed red line) and in the cross-correlation of real and redshift space 
(solid black line).
Note how the BAO peak in redshift or cross-space
is enhanced quite a bit compared to that in real space.
We compare these simulation results with analytic predictions (lines without dots;
blue dotted for real space, red dashed for redshift space and black solid for cross-space) which
combine linear theory on large scales with velocity dispersion on small scales
(see paper I for an extensive explanation of the model).
Note how the analytic model roughly matches the simulation data.
However, despite the similarities in the shape, the amplitude of the LOS baryon peak 
seems slightly larger in the simulations than in the analytic model. 
It is not entirely clear to us how significant the difference is, but
it should be noted that non-linear redshift space distortions have
yet to be properly explored and modeled in the LOS direction \cite{scoccimarro}. 
We will see a similar tendency in the actual LRG galaxies but the measurement
errors are large.
Ultimately, what we need is something like a LOS integral of $\xi_{mm}$ (see Eq. [\ref{eq:xigmu3}]).
We find that the difference between integrating 
$\xi_{mm}(\pi,\sigma)$ in real space, redshift space or in the cross space 
is not very significant.

Next, we repeat the same exercise for $\xi_{gm}$, using halos
in the simulations as a proxy for LRGs. The conclusions are quite
similar to those for $\xi_{mm}$. In particular, we find that
as far as the LOS integral of $\xi_{gm}$ is concerned
(Eq. [\ref{eq:xigmu3}]), it actually does not
matter much whether we use $\xi_{gm}$ in real space, redshift space
or cross-space. What distinguishes $\xi_{gm}$ from $\xi_{mm}$ is
the galaxy bias which is about $2$ on large scales but rise to
about $3 - 4$ on small scales. As emphasized in \S \ref{sec:nonstandard},
the nonlinear galaxy bias affects $\xi_{g\mu}(\pi, \sigma)$ for a small $\sigma$,
even if $\pi$ is large. This is illustrated in Fig. \ref{fig:xigmu}.
The correct $\xi_{g\mu}$ using a nonlinear galaxy bias is larger
than the one using a linear galaxy bias by about a factor of $1.5$.
Note also how the $\pi$ dependence is close to linear, supporting the approximations
made in Eq. (\ref{eq:xigmu}).

Finally, we have also studied the effect of the chosen value for $\chi_2$ in the integral of Eq.\ref{eq:xigmu3}. 
The real magnification should be a weighting of integrals at the different redshifts (or $\chi_2$) of the survey. 
Each redshift bin is weighted depending on the number of galaxies and the volume under this $\chi_2$. 
The exact result is almost identical to taking $\chi_2$ to be at the median redshift of the survey.

\end{document}